\documentclass[final,5p,times,twocolumn]{elsarticle}

\usepackage{amsmath}
\usepackage{bm}
\usepackage{graphicx}
\usepackage{multirow}
\usepackage{float}
\usepackage{tikz}
\usepackage{placeins}
\usepackage{upgreek}
\biboptions{numbers,sort&compress}

\usepackage[hidelinks]{hyperref}

\begin{document}

\begin{frontmatter}
\title{Microstructural Evolution and Crystallization Behavior of Amorphous Medium‑Entropy Ti–Nb–Zr–Ag Thin Films}
\author[inst1]{Anna Benediktov\' a\corref{cor1}}
\ead{amalinov@ntc.zcu.cz}
\author[inst1]{Lucie Nedv\v{e}dov\'a}
\author[inst1]{Michal Proch\'azka}
\author[inst1]{Zden\v{e}k Jansa}
\author[inst1]{\v{S}t\v{e}p\'anka Jansov\'a}
\author[inst2]{Christopher D. Woodgate}
\author[inst3]{David Redka}
\author[inst4]{Julie B. Staunton}
\author[inst1]{J\'an Min\'ar\corref{cor2}}
\ead{jminar@ntc.zcu.cz}
\address[inst1]{New Technologies Research Centre, University of West Bohemia, Univerzitn\'i 8, 301 00 Pilsen, Czech Republic}
\address[inst2]{H.H. Wills Physics Laboratory, University of Bristol, Royal Fort, Bristol, BS8 1TL, United Kingdom}
\address[inst3]{Laser Center HM, Munich University of Applied Sciences HM, Munich, Germany}
\address[inst4]{Department of Physics, University of Warwick, Coventry, CV4 7AL, United Kingdom}
\cortext[cor1]{Corresponding author}
\cortext[cor2]{Co-corresponding author}
\begin{abstract}
Improving the performance of metallic implants increasingly relies on the development of multifunctional surface modifications that combine structural stability, bioactivity, and prevention of bacterial colonization. Medium-entropy alloys (MEAs) represent a promising approach for such coatings, as their chemical complexity allows the formation of structurally stable matrices with tunable properties. In this study, Ti–Nb–Zr and Ti–Nb–Zr–Ag thin films were deposited by magnetron sputtering and subjected to annealing at temperatures of up to 1100~$^{\circ}$C to evaluate the influence of Ag, added for its antibacterial potential, on structural evolution. The as‑deposited Ag‑free film was fully amorphous, whereas the Ag‑containing film exhibited a predominantly amorphous matrix with finely dispersed crystalline nanoparticles, indicating that Ag promoted early‑stage crystallization. Both films displayed a fine columnar morphology (column diameter $\sim$15~nm) with dome‑like protrusions, a hierarchical surface structure favorable for protein adhesion. Upon annealing, the Ag‑free film recrystallized into a granular, loosely packed morphology, while the Ag‑containing film retained a compact structure, demonstrating the stabilizing role of Ag. These findings underscore the potential of Ag‑containing amorphous MEAs for forming multifunctional coatings with enhanced thermal stability, antibacterial functionality, and biointerface‑relevant surface features for advanced biomedical applications.
\end{abstract}
\end{frontmatter}

\section{Introduction}
Load‑bearing biomedical implants require materials with excellent mechanical strength, long‑term stability, and reliable biocompatibility. Although conventional metallic biomaterials such as stainless steels, Co‑Cr alloys, and Ti‑6Al‑4V are widely used in clinical practice~\cite{Niinomi}, they still show limitations when exposed to physiological environments. In addition to concerns about corrosion resistance and the release of potentially cytotoxic or allergenic ions (e.g., Fe, Cr, Ni, Co, Al, V)~\cite{Hallab, Jacobs}, these materials generally lack intrinsic antibacterial properties~\cite{Campoccia} and typically require surface modification to support effective osteointegration~\cite{Anselme, Stich}.

Applying a thin, chemically stable surface coating via magnetron sputtering represents an effective strategy to mitigate these issues. Such coatings can inhibit corrosion processes at the metal surface by limiting the formation of harmful products, while also acting as partial diffusion barriers~\cite{Saad, Padamata}. By modifying only the surface, they allow implants to maintain their mechanical integrity while benefiting from improved surface performance.

Beyond acting as a passive barrier, the coating’s morphology can be tailored to promote bioactivity. Features such as controlled porosity and surface roughness support early protein adsorption and subsequent bone cell attachment. Additionally, stable oxide phases can enhance bioactivity by providing a chemically favorable surface for cellular responses~\cite{Chopra, Anwar}. Thus, such coatings can significantly contribute to successful implant integration.

Incorporating Ag into the coating may provide intrinsic antibacterial properties that help reduce the risk of postoperative infections, a major cause of implant failure~\cite{Savvidou}. While antibiotic‑based strategies are still widely used, increasing bacterial resistance highlights the need for alternative approaches with inherent antimicrobial activity. In addition, silver may influence the thermodynamic stability and crystallization behavior of the coating during deposition or thermal treatment, potentially enhancing its mechanical performance~\cite{Rashid}, and has been reported to improve corrosion resistance in similar Ag‑containing systems~\cite{Zhang, Rosalbino}.

We selected the alloy compositions studied in this work using a parametric approach typical for high‑ and medium‑entropy alloy (HEA, MEA) design strategies~\cite{Yeh}. Elements were chosen from a subset of well‑established biocompatible metals~\cite{Niinomi} to identify a combination likely to form a solid solution with Ag while ensuring high corrosion resistance, stable oxide formation, and a favorable biological response. Ti, Nb and Zr were chosen based on their calculated atomic size mismatch ($\delta$), electronegativity difference ($\Delta\chi$), mixing enthalpy ($\Delta H_{\rm mix}$), and the thermodynamic stability parameter $\Omega$ in combination with Ag. Given the four principal elements, the configurational entropy of mixing of the system ($\Delta S_{\rm mix}$) falls within the range typical of MEAs~\cite{Ma}.

Such multi-principal element alloys offer a promising pathway toward materials with enhanced stability, strength, and resistance to degradation in physiological environments~\cite{Senkov}, all desirable for biomedical coatings. In principle, magnetron sputtering enables precise control of film thickness and microstructure together with uniform composition. However, the non-equilibrium nature of the process and the typically low substrate temperatures, where growth is governed more by kinetics than thermodynamics, often drive these complex alloys into metastable or amorphous states~\cite{Padamata}. The parameter-based design approach, originally developed for the formation of bulk alloys under near-equilibrium conditions, therefore serves primarily as qualitative guidance rather than a precise predictor of the as-deposited phase constitution. 

Such deposition conditions open up the possibility of forming amorphous medium-entropy alloys, which have recently attracted interest for their exceptional mechanical properties, such as high hardness and wear resistance, accompanied by excellent corrosion resistance~\cite{Zhang2, Garah}. Amorphous MEAs are particularly attractive for biomedical applications, as they eliminate grain boundaries, common initiation sites for corrosion or ion release. These alloys may also contain nanoscale crystalline domains or intermetallic phases, which can contribute to improved mechanical performance~\cite{Amorphous}. Nevertheless, they remain underexplored, especially in systems combining biocompatible elements with Ag for multifunctional surface coatings.

Despite their great potential, only a few studies have examined Ti–Nb–Zr‑based coatings, and, to our knowledge, just one report exists on quaternary Ti–Nb–Zr–Ag films~\cite{Liu}, with their possible microstructural states and related functional properties insufficiently characterized. In this study, therefore, we present a detailed investigation of how the addition of Ag affects the growth and structure of magnetron‑sputtered Ti–Nb–Zr‑based thin films, as well as their crystallization behavior during controlled annealing. Using a combination of scanning and transmission electron microscopy and X‑ray photoelectron spectroscopy, we aim to clarify the influence of Ag on phase formation, growth mechanisms, crystallization, and surface chemistry in order to optimize the coating's properties for biomedical applications. The mechanical and corrosion behavior of as-deposited thin films, as well as microstructure evolution during annealing, will be addressed in subsequent studies.

The remainder of this paper is organized as follows. Section~\ref{sec:design_and_experiment} describes the design ideas and experimental procedures, including magnetron sputtering, scanning and transmission electron microscopy, and X-ray photoelectron spectroscopy. Section~\ref{sec:results_and_discussion} then presents and discusses the results, focusing on the morphology, microstructure, and chemical characterization of as-deposited and annealed Ti–Nb–Zr and Ti–Nb–Zr–Ag films. Finally, in Section~\ref{sec:summary_and_conclusions}, the results are summarized and appropriate conclusions are drawn. 

\section{Design ideas and experimental procedures}
\label{sec:design_and_experiment}

\subsection{Design of thin films}

\begin{table*}[t]
\centering
\caption{Calculated thermodynamic parameters for the Ti–Nb–Zr(–Ag) alloys.}
\label{parameters}
\begin{tabular}{|l|c|c|c|c|c|}
\hline
{Alloy} & ${\Delta S_{\rm mix}}$~(J$\cdot$mol$^{-1}$\,K$^{-1}$) & ${\delta}~(\%)$ & ${\Delta H_{\rm mix}}$~(kJ$\cdot$mol$^{-1}$) & ${\Delta \chi}~(\%)$ & ${\Omega}$ \\
\hline
Ti$_{0.78}$Nb$_{1.08}$Zr$_{1.14}$ & 9.03 & 4.36 & 2.94 & 8.17 & 7.07 \\
\hline
Ti$_{0.64}$Nb$_{1.44}$Zr$_{1.40}$Ag$_{0.52}$ & 10.76 & 4.51 & 1.67 & 12.42 & 14.21 \\
\hline
\end{tabular}
\end{table*}

To evaluate the phase stability of the Ti–Nb–Zr–Ag system, a parametric approach based on empirical design criteria for multi-principal element alloys was applied. The following parameters were calculated from the atomic fractions and elemental properties of the selected metals:

\begin{itemize}
\item \textbf{Entropy of mixing}

The configurational entropy, typically assumed to be the dominant contribution to the total entropy of mixing $\Delta S_{\text{mix}}$, is given by
\begin{equation}\label{con}
\Delta S_{\rm conf} = -R \sum_{i=1}^n c_i \log{c_i} \, ,
\end{equation}
where $R$ is the ideal gas constant and $c_i$ is the atomic fraction of the $i$-th of $n$ components. Vibrational ($\Delta S_{\rm vib}$), magnetic ($\Delta S_{\rm mag}$), and electronic ($\Delta S_{\rm elec}$) contributions are often neglected in a first-order approximation. Larger configurational entropy values typically promote the formation of disordered solid solutions. Consequently, alloys can be roughly classified as low-, medium- and high-entropy for $\Delta S_{\rm conf} < 1.0\, R$, $1.0\, R \le \Delta S_{\rm conf} < 1.5\, R$, and $\Delta S_{\rm conf} \ge 1.5\, R$, respectively~\cite{Bolar}. 
In absolute units, these thresholds correspond to
$\Delta S_{\rm conf} < 8.3$, $8.3 \le \Delta S_{\rm conf} < 12.5$, and $\Delta S_{\rm conf} \ge 12.5~\mathrm{J\cdot mol^{-1}\,K^{-1}}$.

\item \textbf{Atomic size mismatch}
\begin{equation}\label{delta}
\delta = 100\cdot\sqrt{\sum_{i=1}^n c_i\bigg(1-\frac{r_i}{\bar{r}}\bigg)^2}\, ,      
\end{equation}
where $r_i$ is the atomic radius of the $i$-th constituent element and $\bar{r}$ is the weighted average radius ${\bar r}=\sum_{i=1}^n c_i r_i$. Solid solutions are commonly reported to form for $\delta \leq 6.6$~\%~\cite{Yang2012}.

\item \textbf{Enthalpy of mixing}
\begin{equation}\label{H}
\Delta H_{\rm mix} = \sum_{i<j}^n 4\Delta H_{\rm mix}^{\rm AB}c_ic_j\, ,   
\end{equation}
where $\Delta H_{\rm mix}^{\rm AB}$ are binary mixing enthalpies (e.g., from Miedema's semiempirical model~\cite{Miedema}). Solid solutions typically form for $-22~\mathrm{kJ/mol} \leq \Delta H_{\rm mix} \leq 7~\mathrm{kJ/mol}$~\cite{guo}.

\item \textbf{Electronegativity mismatch}
\begin{equation}
\Delta\chi = 100\cdot\sqrt{\sum_{i=1}^{n} c_i \left( 1-\frac{\chi_i}{\bar{\chi}}\right)^2}\, ,
\end{equation}
where $\chi_i$ is the Pauling electronegativity of the $i$-th element and $\bar{\chi}$ is the weighted average ${\bar \chi}=\sum_{i=1}^n c_i \chi_i$. A solid solution is predicted to form when $\Delta \chi < 6\,\%$~\cite{Caramarin}.

\item \textbf{Thermodynamic stability parameter} 
\begin{equation}\label{omega}
\Omega= \frac{T_{\rm m}\Delta S_{\rm mix}}{|\Delta H_{\rm mix}|}\, ,         
\end{equation}
where $T_m$ is the weighted average melting temperature of the constituent elements, $T_m =\sum_{i=1}^n c_i T_{m_i}$. A critical value of $\Omega \approx 1$ separates intermetallic-dominated systems from solid-solution forming ones; for $\Omega > 1.1$, entropy dominates over enthalpy and a solid solution is favored~\cite{Yang2012}.
\end{itemize}

The values of the above-mentioned parameters for the studied thin films with compositions Ti (26 at.\%), Nb (36 at.\%), Zr (38 at.\%) and Ti (16 at.\%), Nb (36 at.\%), Zr (35 at.\%), Ag (13 at.\%), are listed in Table~\ref{parameters}. Based on the configurational entropy, both compositions fall into the MEA category. All calculated thermodynamic parameters are within the ranges often associated with solid-solution formation: $\delta$ was calculated using metallic atomic radii of Ti (147~pm), Nb (146~pm), Zr (160~pm), and Ag (144~pm), yielding a low $\delta$ value indicative of good atomic-size compatibility. The enthalpy of mixing, calculated using binary mixing enthalpies $\Delta H_{\rm mix}^{\rm AB}$ reported by Hussein et al.~\cite{Hussein}, is relatively close to zero and the $\Omega$ parameter exceeds the commonly used threshold of 1.1, both suggesting a dominant entropic contribution. The only exception is the electronegativity mismatch $\Delta \chi$, which is higher and may indicate a tendency toward the formation of ordered phases.

\subsection{Experimental procedures and characterizations}
\subsubsection{Magnetron Sputtering}

All Ti–Nb–Zr and Ti–Nb–Zr–Ag thin films investigated in this work were deposited using a BOC Edwards TF600 coating system equipped with both radio-frequency (RF) and direct-current (DC) magnetron sputtering sources. Single-crystal Si (100) wafers, chemically cleaned prior to deposition, served as substrates. Before each deposition, the vacuum chamber was evacuated to a base pressure of $2\cdot 10^{-4}$~Pa using a combination of a dry rotary pump and a turbomolecular pump.

The ternary Ti–Nb–Zr films were sputtered from an equimolar Ti–Nb–Zr target (purity
99.99~\%) mounted on an RF magnetron operating at a frequency of 13.56~MHz. For the
quaternary Ti–Nb–Zr–Ag films, co-sputtering was employed using the equimolar Ti–Nb–Zr target on the RF magnetron together with a high-purity Ag target (purity 99.99~\%) powered by DC.

All depositions were performed in high-purity argon at a working pressure of 0.8~Pa.
Power densities referenced to the measured racetrack were $\sim$21.5~W$\cdot$cm$^{-2}$ for the RF-driven Ti–Nb–Zr target and $\sim$0.72~W$\cdot$cm$^{-2}$ for the DC-powered Ag target. No external heating or substrate bias was applied, leaving the substrates at floating potential. The deposition process was adjusted to obtain a film thickness of $\sim$350~nm (sufficient for the subsequent characterization), corresponding to deposition rates of 8.8~nm/min for Ti–Nb–Zr and 10.0~nm/min for Ti–Nb–Zr–Ag.

The as-deposited films were subsequently annealed in a high-temperature vacuum chamber (HTK Anton Paar 1200), maintained at a vacuum of approximately 10$^{-3}$~Pa, at temperatures up to 1100~$^{\circ}$C. The annealing was performed in a graded manner with intermediate stabilization at selected temperatures to ensure thermal equilibrium during heating. 

\subsubsection{Scanning and Transmission Electron Microscopy (SEM), (TEM)}

The surface and cross-sectional morphology of the films was examined using two scanning electron microscopes: JEOL JSM-IT710HR and JEOL JSM-7600F. Cross-sectional SEM specimens were prepared by fracturing the coated Si substrates after immersion in liquid nitrogen. The fracture surfaces were observed at an angle of 45$^{\circ}$.

Cross-sectional TEM specimens of as-deposited films were prepared using a JEOL IB-09060CIS cryo ion slicer. Cross-sections of annealed films were prepared using a ZEISS AURIGA focused ion beam (FIB) system. To protect the film surface during the preparation of the FIB lamella, a thin platinum layer was first deposited by electron-beam deposition. 

TEM and scanning TEM (STEM) analyses were performed on a JEOL JEM-2200FS transmission electron microscope operated at 200 kV. Phase identification was based on the combination of selected-area electron diffraction (SAED), high-angle annular dark-field (HAADF) imaging, energy-dispersive X-ray spectroscopy (EDX) in STEM mode (elemental mapping and point analyses), TEM, and high-resolution TEM (HRTEM) imaging with fast Fourier transform (FFT) and inverse FFT reconstruction performed in Digital Micrograph (Gatan, USA). Experimental FFT and SAED patterns were compared with simulated diffraction patterns, and simulated HRTEM images were generated using the multislice method. All simulations were carried out in JEMS (P. Stadelmann, Jongny, Switzerland).

\subsubsection{X-Ray Photoelectron Spectroscopy (XPS)}

X-ray photoelectron spectroscopy (XPS) analysis of Ti--Nb--Zr and Ti--Nb--Zr--Ag thin films was performed in an ultra-high vacuum (UHV) chamber with a base pressure of $\leq 3 \cdot 10^{-8}$~Pa, using a Phoibos 150 hemispherical analyzer (SPECS Surface Nano Analysis GmbH) and a non-monochromatic XR 50 X-ray source operated with the Mg K$\alpha$ line (1253.6~eV). Core-level spectra were acquired over 20 scans with an energy step size of 0.05~eV and a pass energy of 30~eV. The samples were mounted on a molybdenum sample holder using conductive Ag paste. Measurements were performed both in the as-introduced state and after Ar$^+$ ion sputtering (30 minutes per cycle at 1 kV).

\section{Results and Discussion}
\label{sec:results_and_discussion}

\subsection{Microstructure and morphology of as-deposited films}

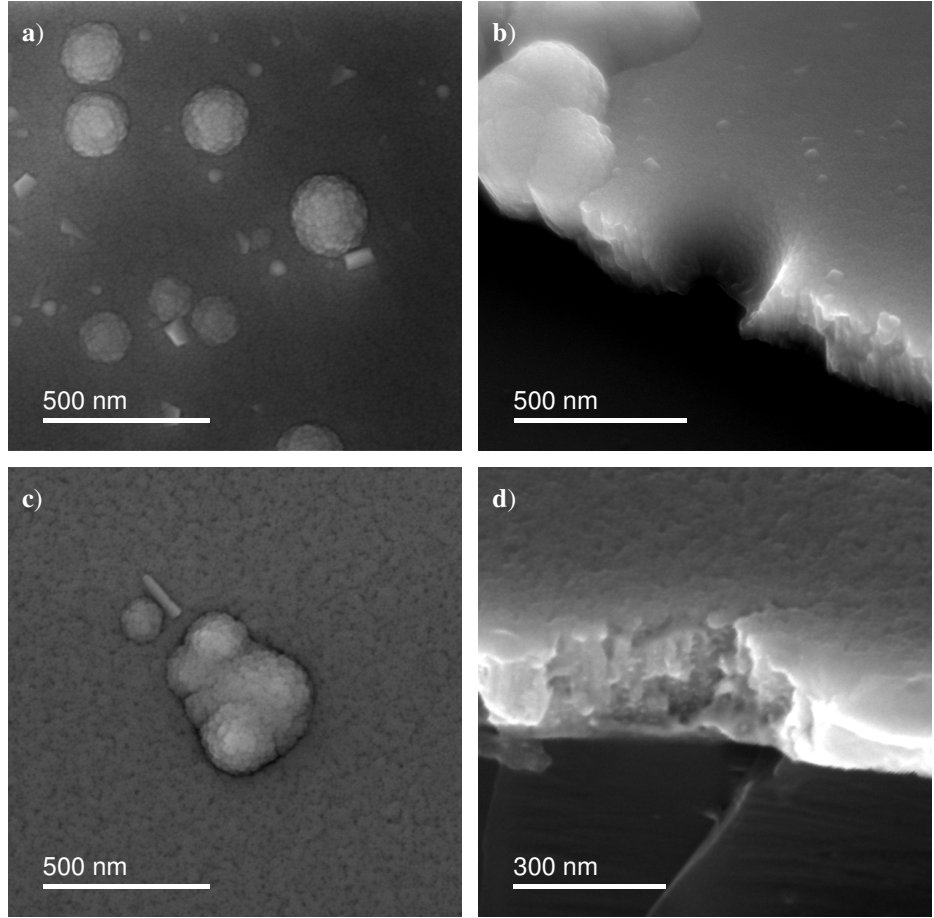
\begin{figure*}[t]
\centering
\includegraphics[width=0.325\textwidth]{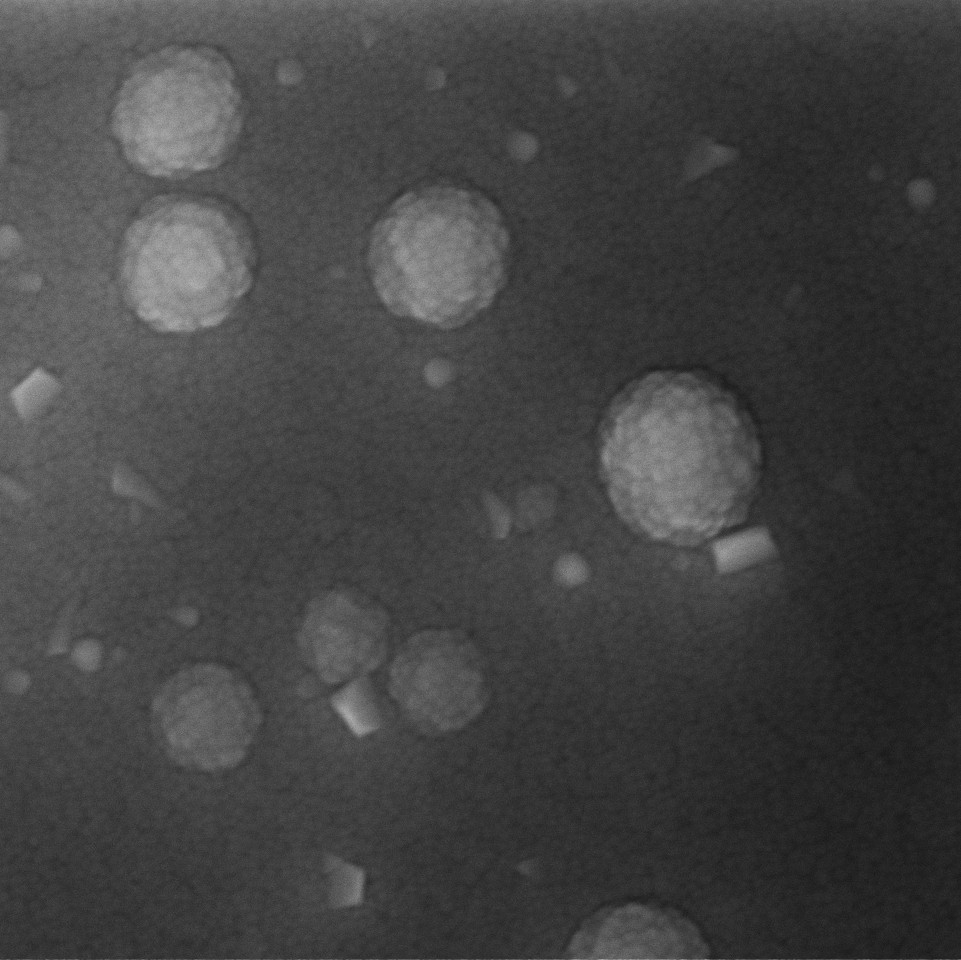}\put(-0.3\textwidth,0.02\textwidth){\color{white}\rule{0.1195\textwidth}{0.0035\textwidth}}\hspace{0.0125\textwidth}%
\put(-0.3\textwidth,0.03\textwidth){\sf\color{white}500 nm}\hspace{0.0125\textwidth}%
\put(-0.315\textwidth,0.296\textwidth){\color{white}{\normalsize\textbf a)}}\hspace{0.0125\textwidth}%
\includegraphics[width=0.325\textwidth]{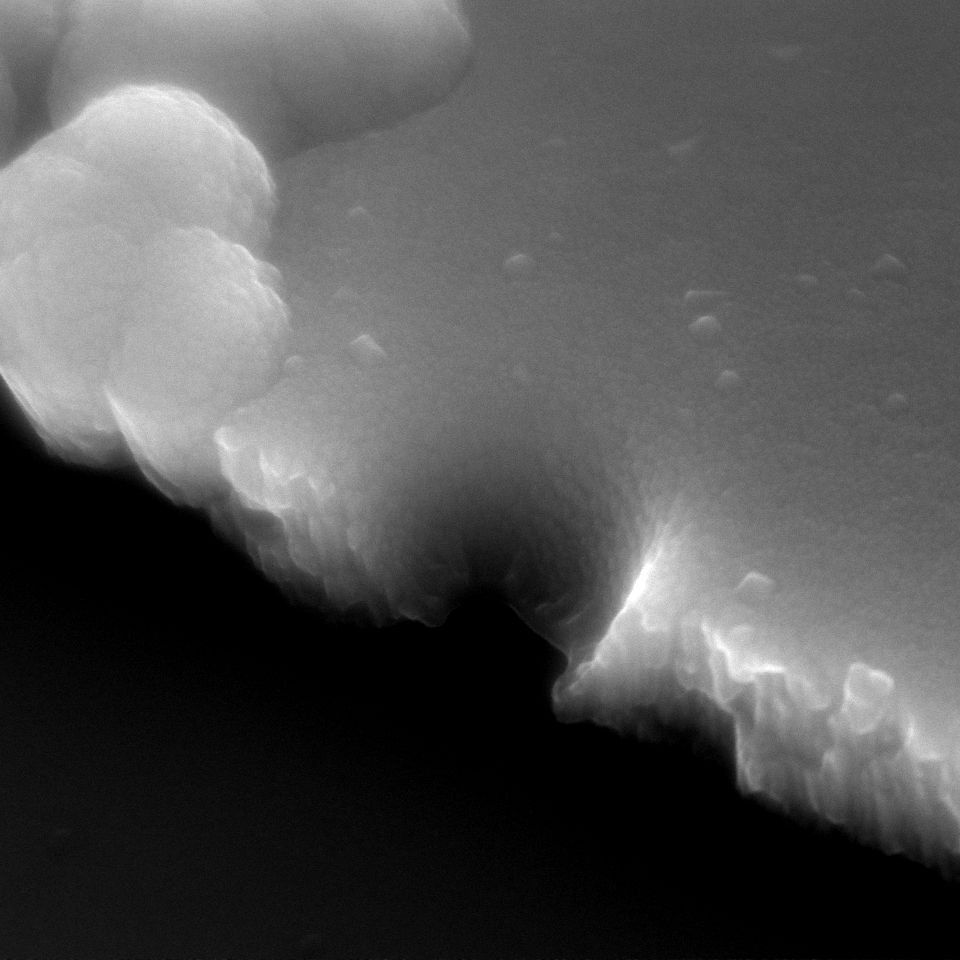}\put(-0.3\textwidth,0.02\textwidth){\color{white}\rule{0.124\textwidth}{0.0035\textwidth}}
\vspace{0.01\textwidth}%
\put(-0.3\textwidth,0.03\textwidth){\sf\color{white}500 nm}%
\put(-0.315\textwidth,0.296\textwidth){\color{white}{\normalsize\textbf b)}}%

\includegraphics[width=0.325\textwidth]{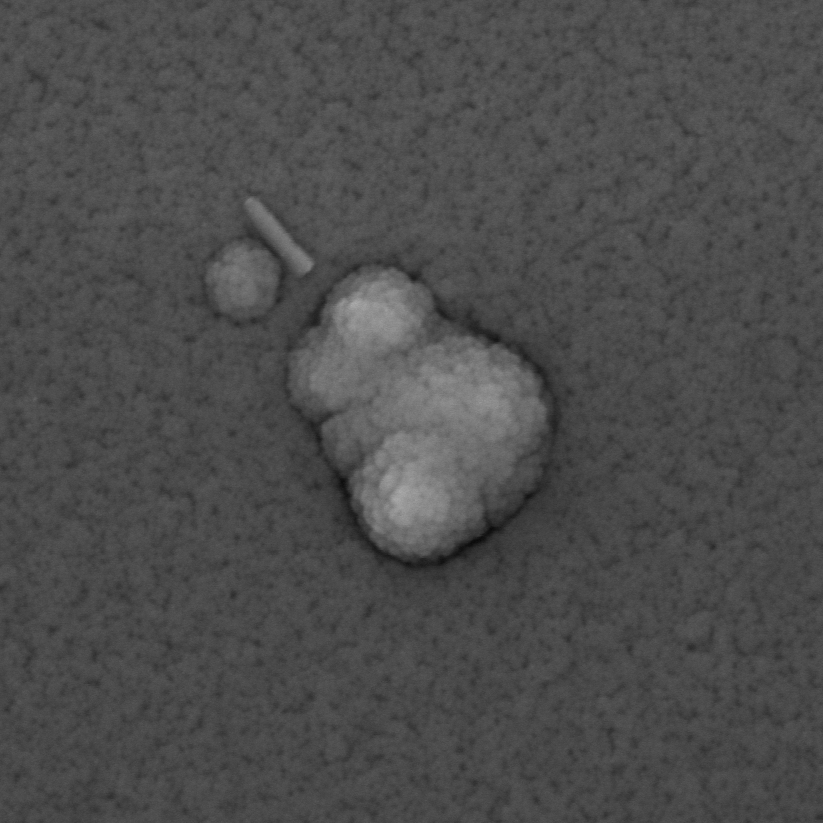}\put(-0.3\textwidth,0.02\textwidth){\color{white}\rule{0.1195\textwidth}{0.0035\textwidth}}\hspace{0.0125\textwidth}%
\put(-0.3\textwidth,0.03\textwidth){\sf\color{white}500 nm}\hspace{0.0125\textwidth}%
\put(-0.315\textwidth,0.296\textwidth){\color{white}{\normalsize\textbf c)}}\hspace{0.0125\textwidth}%
\includegraphics[width=0.325\textwidth]{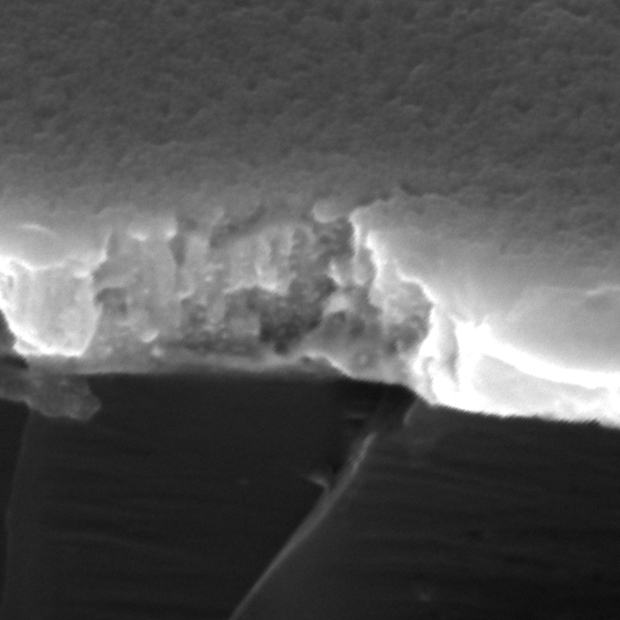}\put(-0.3\textwidth,0.02\textwidth){\color{white}\rule{0.1092\textwidth}{0.0035\textwidth}}
\put(-0.3\textwidth,0.03\textwidth){\sf\color{white}300 nm}%
\put(-0.315\textwidth,0.296\textwidth){\color{white}{\normalsize\textbf d)}}%
\caption{\label{SEMasdep} \small  SEM micrographs of as-deposited Ti–Nb–Zr (a,b) and Ti–Nb–Zr–Ag (c,d) thin films. (a,c) plan-view; (b,d) cross-section 45$^{\circ}$ tilt.} 
\end{figure*}

SEM analysis of the as-deposited Ti–Nb–Zr film revealed that the film was composed of vertical nanocolumns approximately 15~nm in diameter, terminating in rounded caps that collectively formed a smooth, level surface. Above this surface, hemispherical protrusions (\(\sim\)100~nm in diameter) were observed (Fig.~\ref{SEMasdep}a). Cross-sectional observations (Fig.~\ref{SEMasdep}b) showed that these protrusions originated at the substrate–film interface and broadened upward in an inverted-cone shape, indicating heterogeneous nucleation during the initial growth stage. They exhibited the same internal columnar morphology as the surrounding matrix, confirming that they were nodular defects initiated by seed particles on the substrate surface (minor morphological irregularities or contaminants). Here, protrusions occurred unintentionally. However, their area density is tunable by substrate preparation (suppressed by cleaning/smoothing, promoted by deliberate seeding/roughening)~\cite{Panjan}. In addition, small faceted particles (20–40~nm) were observed. Their faceted morphology and surface localization suggest that they may correspond to crystalline surface oxides formed post-deposition, as corroborated by XPS.

\begin{figure*}[t]
\centering
\includegraphics[width=0.325\textwidth]{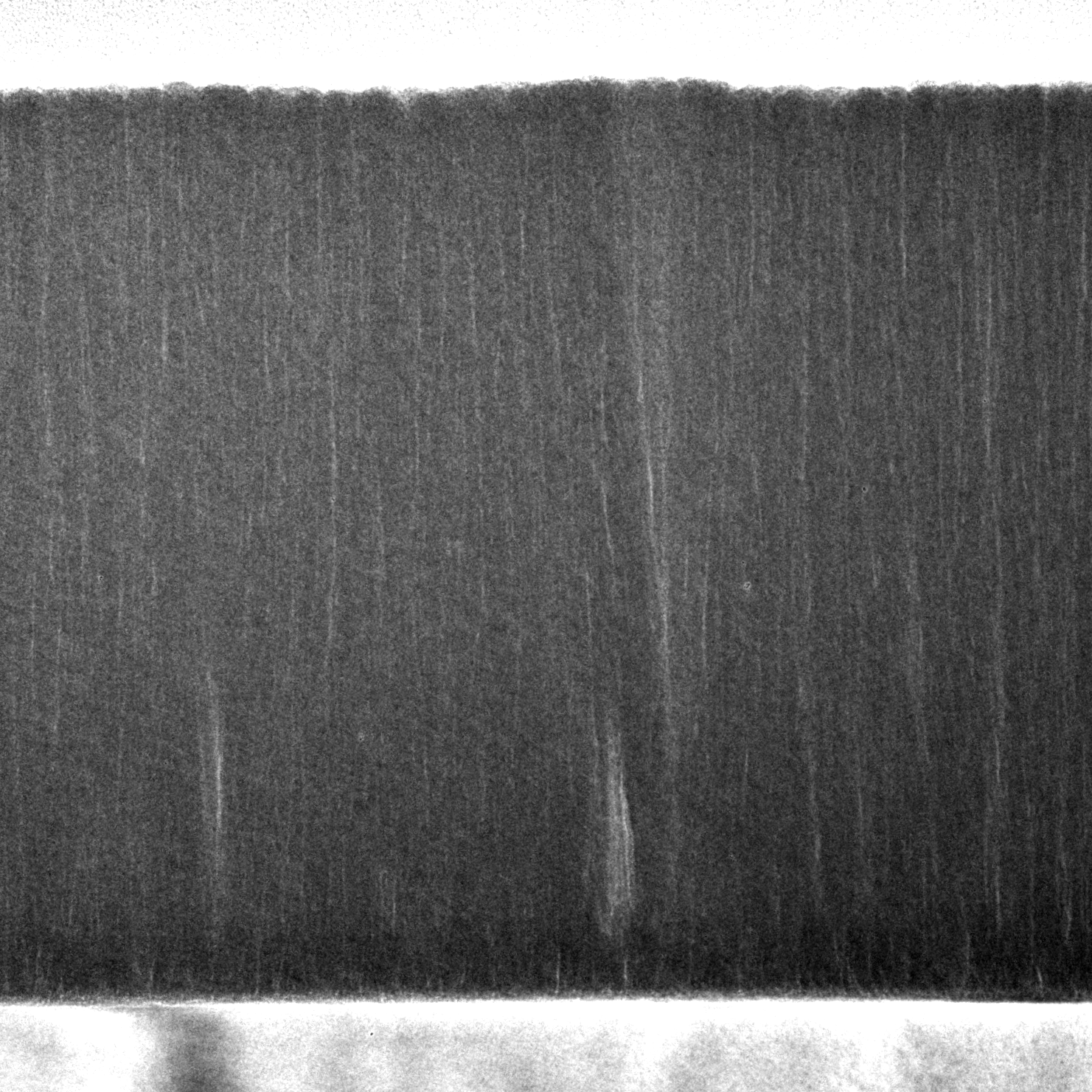}\put(-0.3\textwidth,0.02\textwidth){\color{black}\rule{0.093\textwidth}{0.0035\textwidth}}\hspace{0.0125\textwidth}%
\put(-0.3\textwidth,0.03\textwidth){\sf\color{white}100 nm}\hspace{0.0125\textwidth}%
\put(-0.315\textwidth,0.296\textwidth){\color{black}{\normalsize\textbf a)}}\put(0,0.169\textwidth){\includegraphics[width=0.156\textwidth]{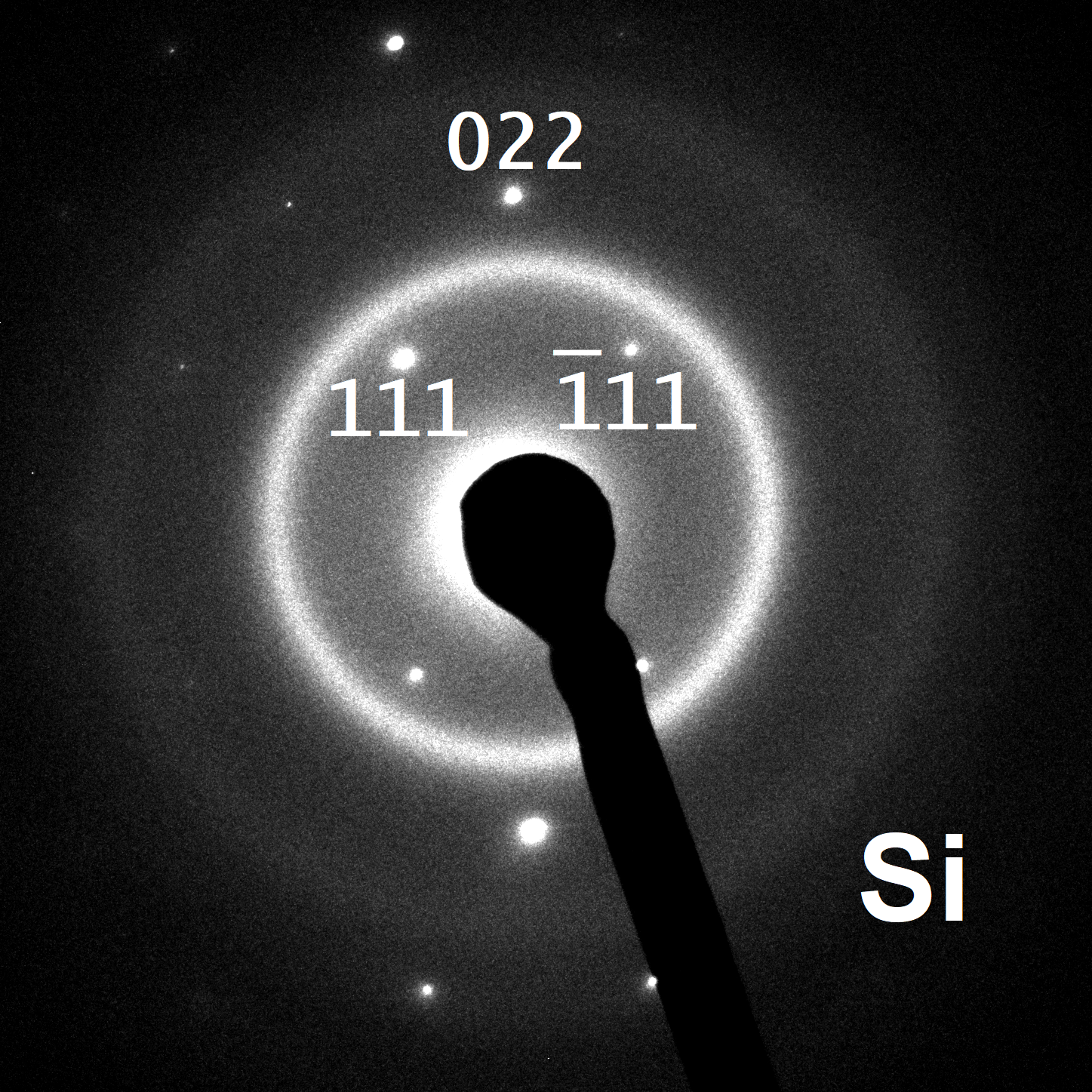}}\hspace{0.0125\textwidth}%
\put(0.01\textwidth,0.299\textwidth){\color{white}{\normalsize\textbf a')}}\hspace{0.0125\textwidth}%
\includegraphics[width=0.156\textwidth]{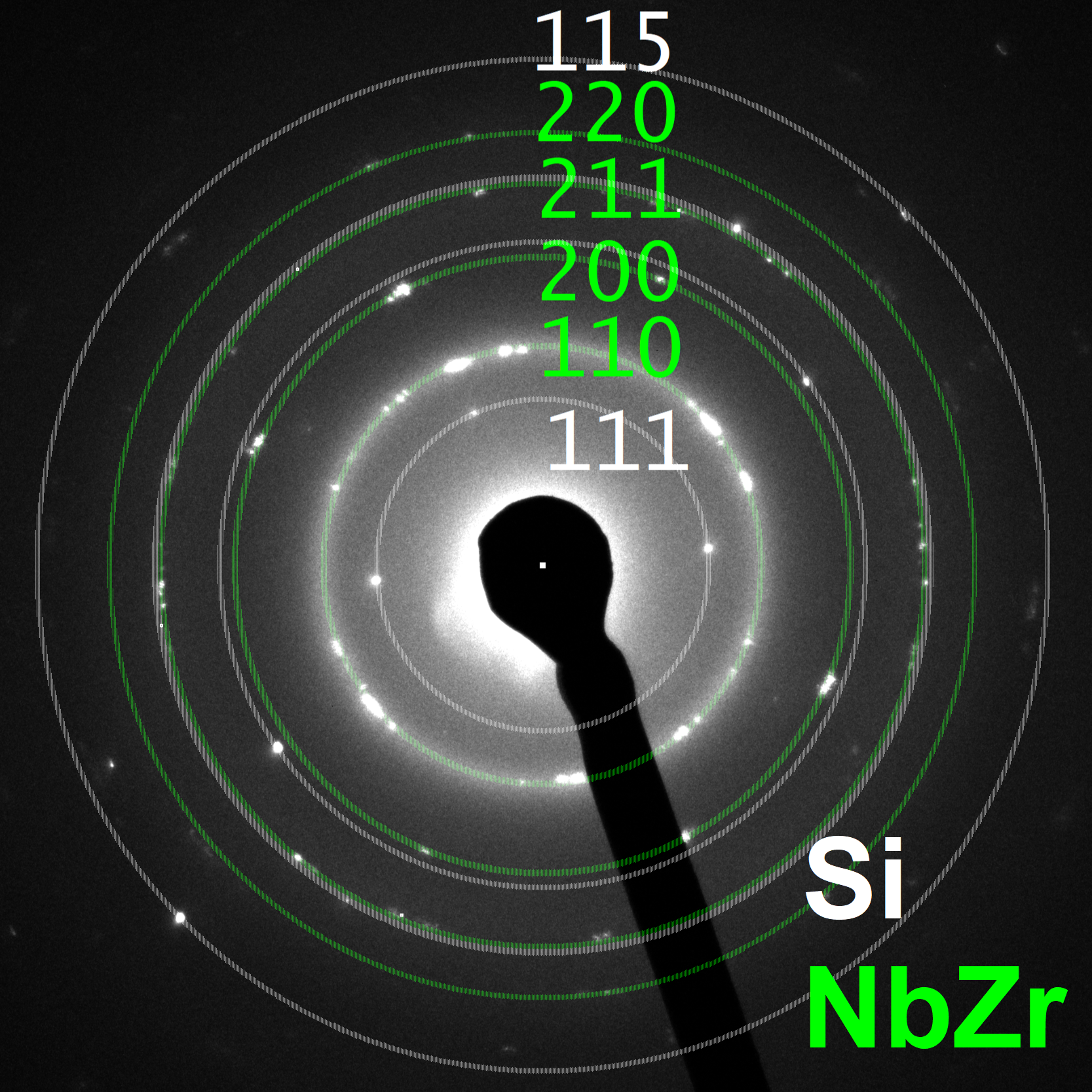}%
\includegraphics[width=0.325\textwidth]{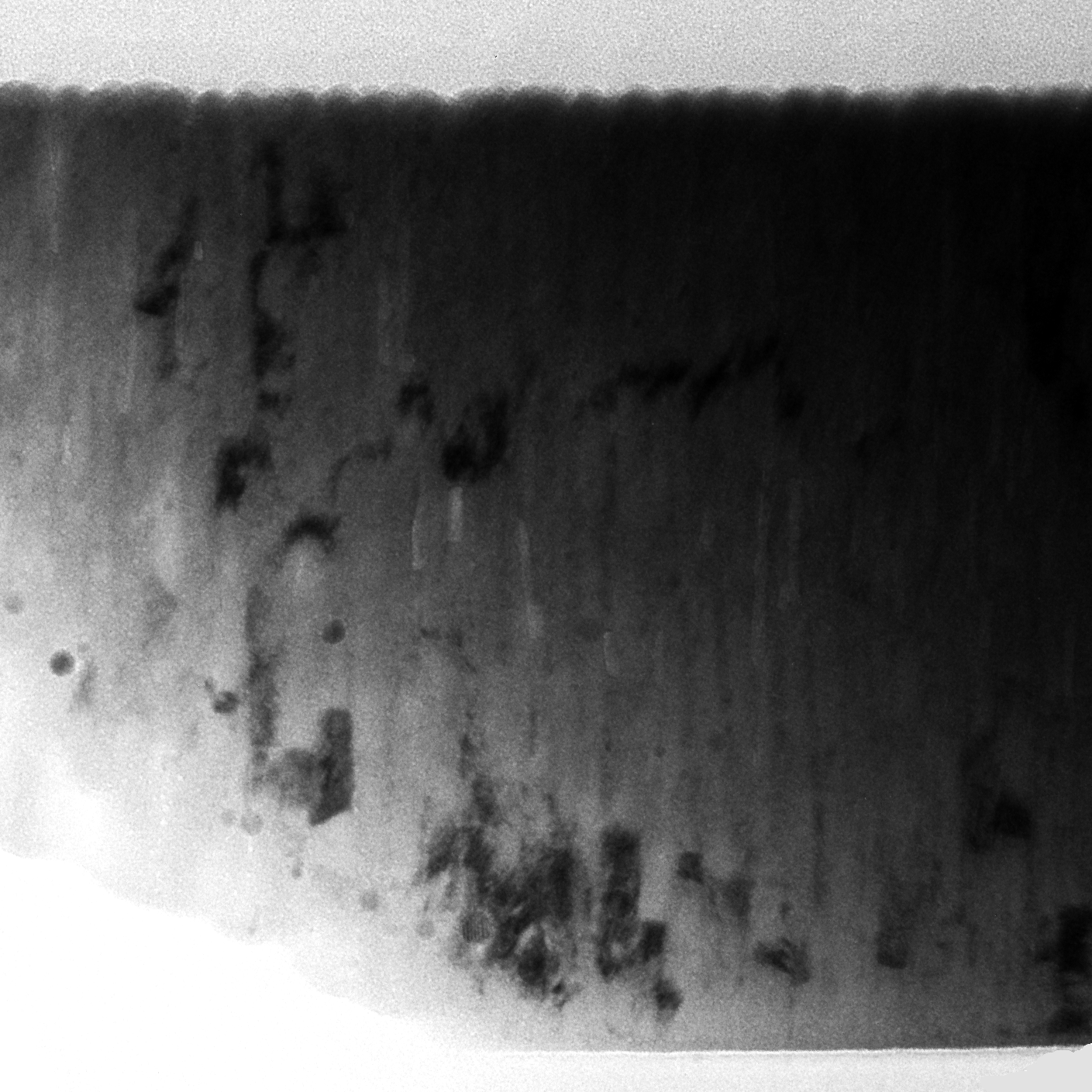}\put(-0.3\textwidth,0.02\textwidth){\color{black}\rule{0.098\textwidth}{0.0035\textwidth}}
\vspace{0.01\textwidth}%
\put(-0.3\textwidth,0.03\textwidth){\sf\color{black}100 nm}%
\put(-0.315\textwidth,0.296\textwidth){\color{white}{\normalsize\textbf b)}}%
\put(-0.472\textwidth,0.131\textwidth){\color{white}{\normalsize\textbf b')}}%

\includegraphics[width=0.325\textwidth]{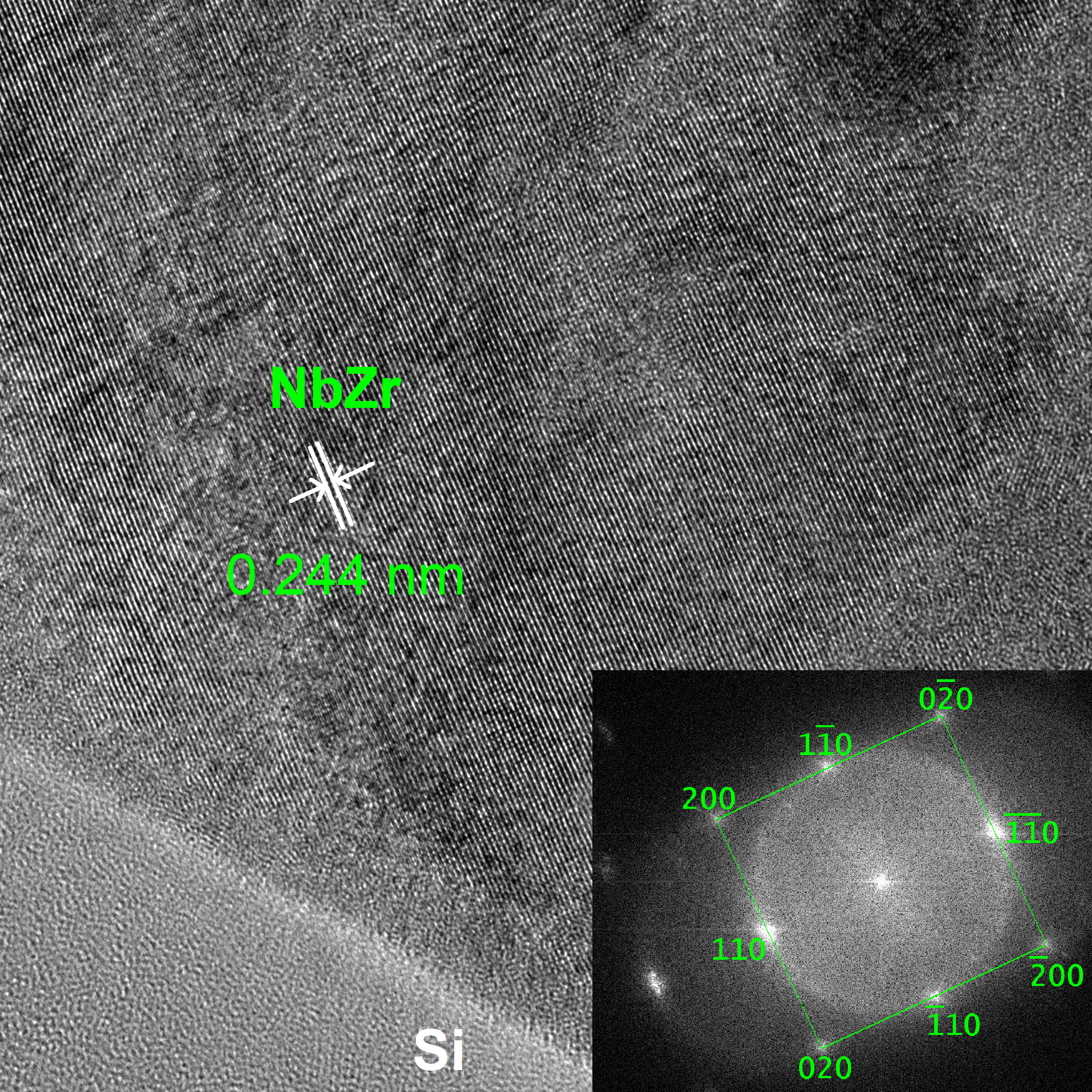}\put(-0.3\textwidth,0.02\textwidth){\color{white}\rule{0.064\textwidth}{0.0035\textwidth}}\hspace{0.0125\textwidth}%
\put(-0.3\textwidth,0.03\textwidth){\sf\color{white}10 nm}\hspace{0.0125\textwidth}%
\put(-0.315\textwidth,0.296\textwidth){\color{white}{\normalsize\textbf c)}}\hspace{0.0125\textwidth}%
\includegraphics[width=0.325\textwidth]{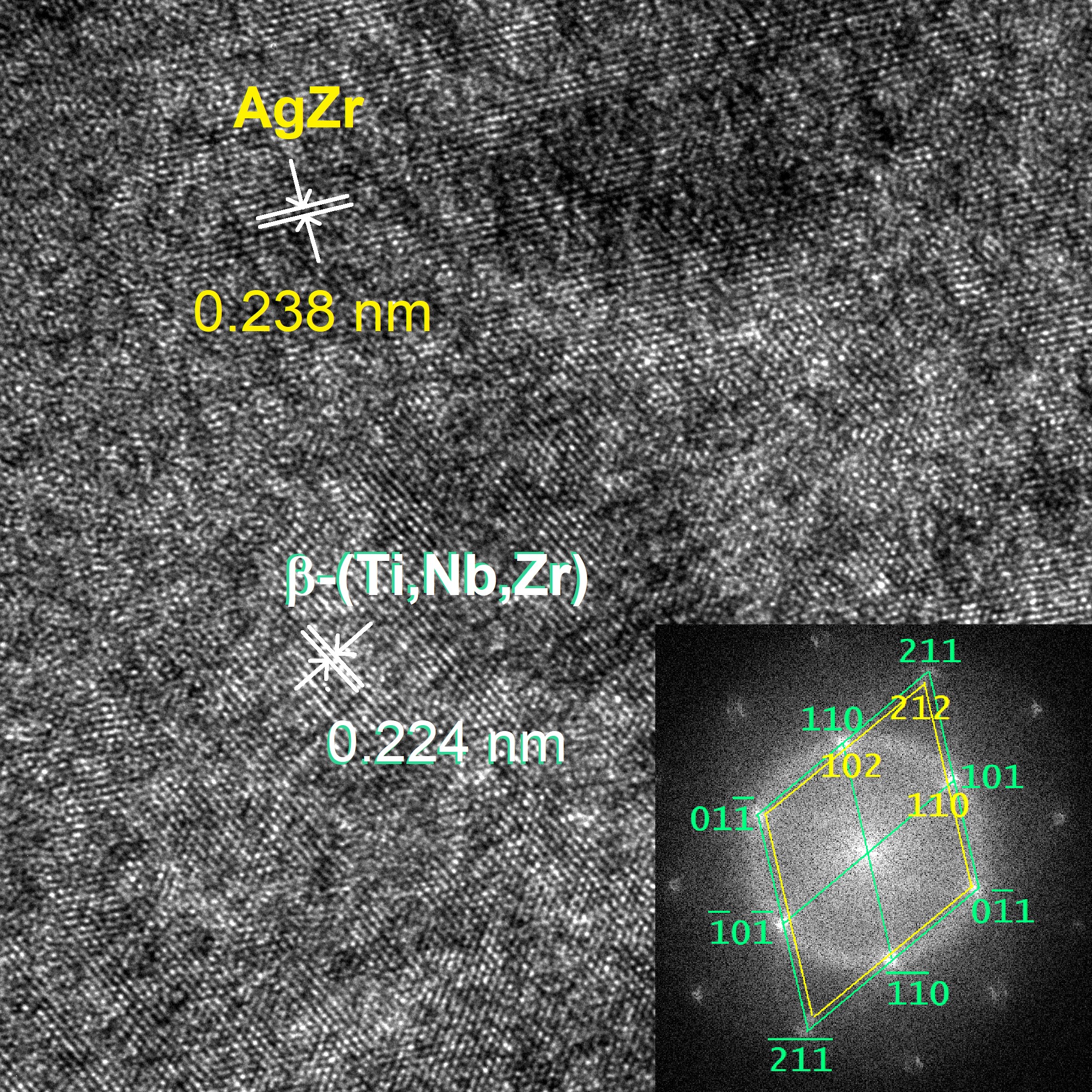}\put(-0.3\textwidth,0.02\textwidth){\color{white}\rule{0.0515
\textwidth}{0.0035\textwidth}}\hspace{0.0125\textwidth}
\put(-0.3\textwidth,0.03\textwidth){\sf\color{white}5 nm}\hspace{0.0125\textwidth}%
\put(-0.315\textwidth,0.296\textwidth){\color{white}{\normalsize\textbf d)}}\hspace{0.0125\textwidth}%
\includegraphics[width=0.325\textwidth]{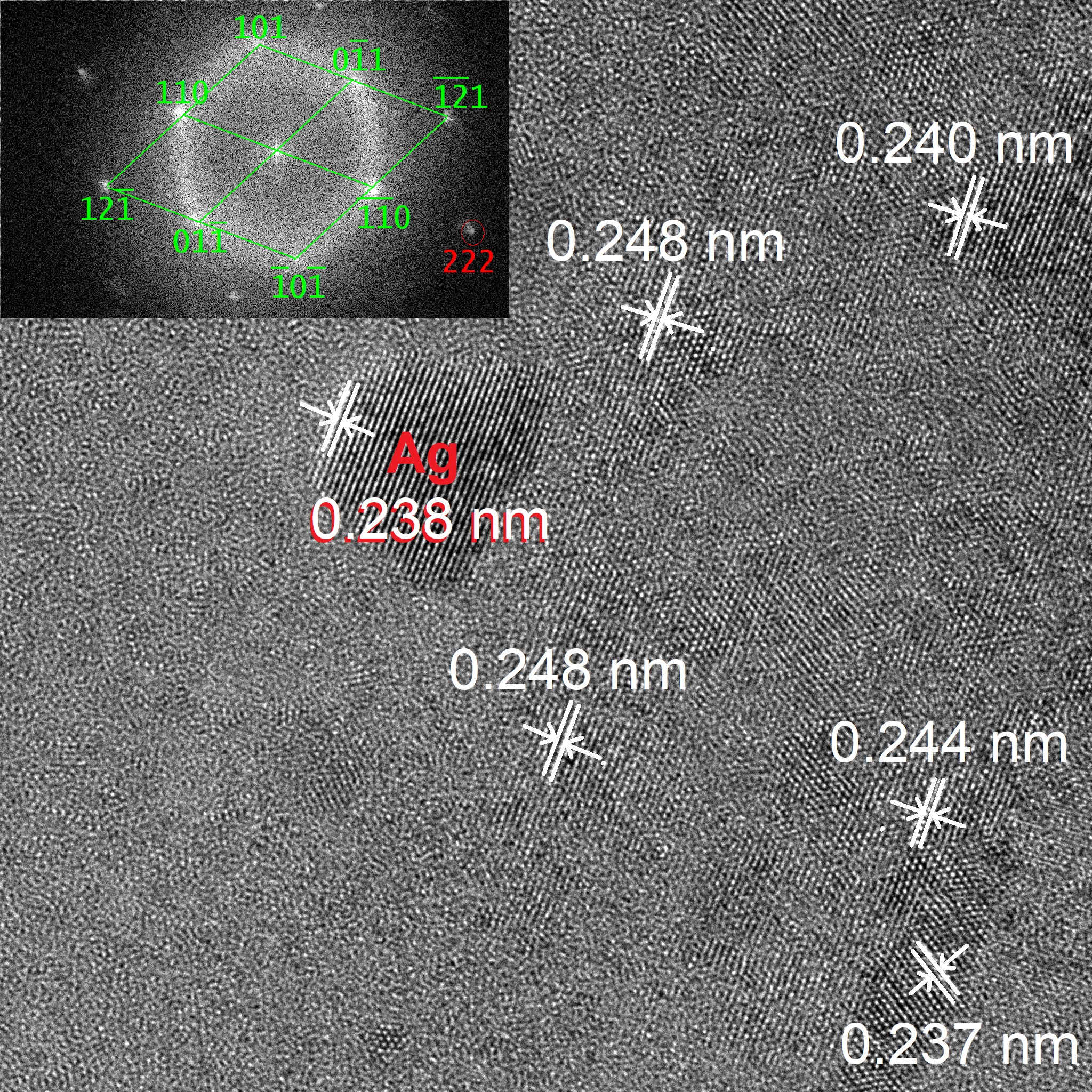}\put(-0.3\textwidth,0.02\textwidth){\color{white}\rule{0.039\textwidth}{0.0035\textwidth}}
\put(-0.3\textwidth,0.03\textwidth){\sf\color{white}5 nm}%
\put(-0.315\textwidth,0.296\textwidth){\color{white}{\normalsize\textbf e)}}%
\caption{\label{TEMasdep} \small Cross-sectional TEM of Ti–Nb–Zr (a,a$^\prime$) and Ti–Nb–Zr–Ag (b–e) thin films. (a) Cross-section overview; (a$^\prime$) SAED with Si [01$\bar{1}$] reflections; (b) Cross-section overview; (c) HRTEM of $\beta$-(Nb,Zr) [001] with $d$-spacing and FFT inset; (d) HRTEM showing AgZr [2$\bar{2}\bar{1}$] and $\beta$-(Ti,Nb,Zr) [$\bar{1}$11], FFT inset; (e) Crystalline seeds with $d$-spacings and Ag nanoparticle, FFT inset.} 
\end{figure*}
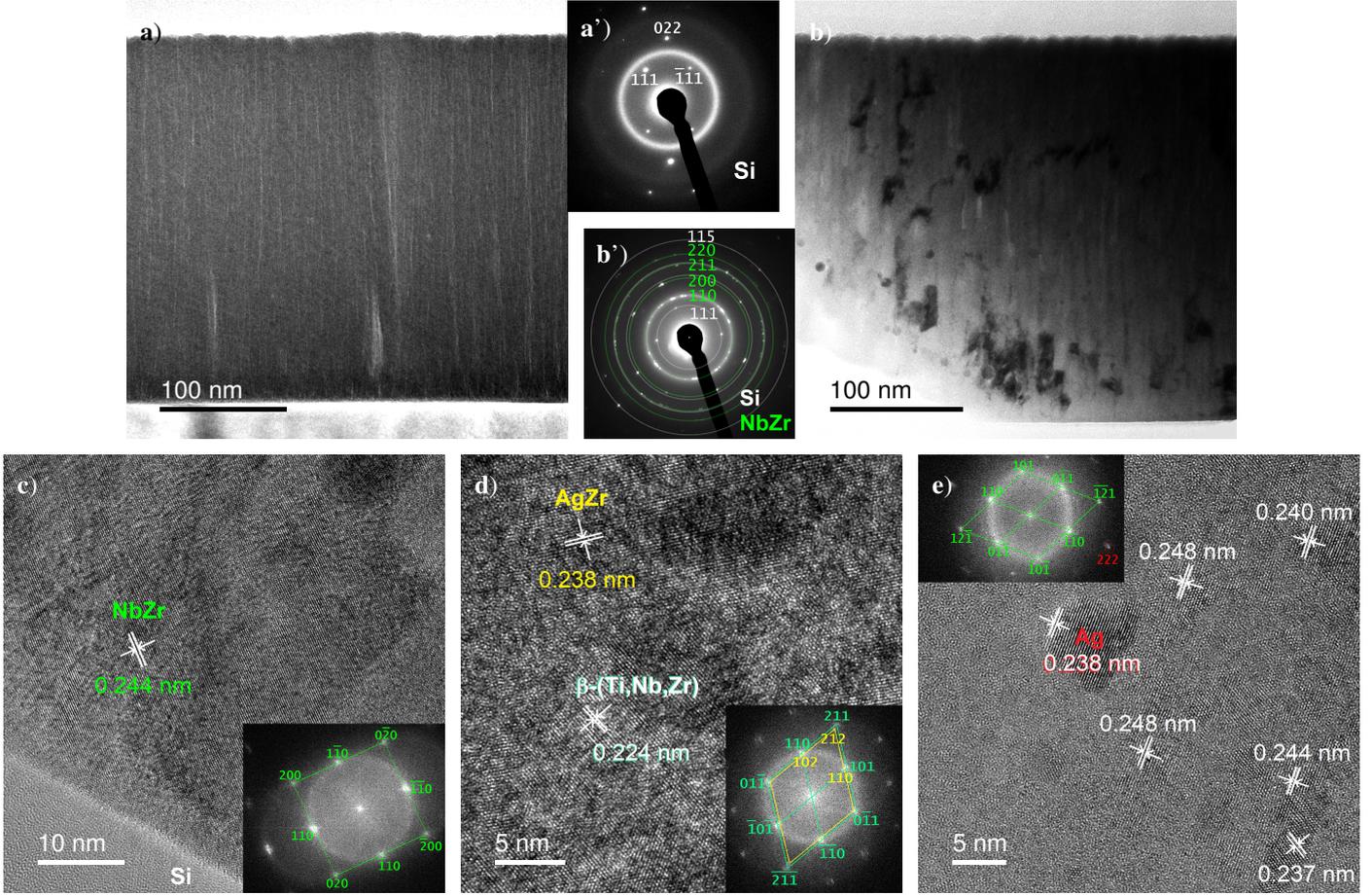

TEM analysis confirmed that the Ti–Nb–Zr film was predominantly amorphous. SAED revealed a single broad diffuse ring (Fig.~\ref{TEMasdep}a), corresponding to an interplanar spacing of approximately 0.25~nm, consistent with the absence of long-range ordering. The film thickness was approximately 293~nm, and its internal structure consisted of vertically aligned nanocolumns, in agreement with SEM observations, suggesting island nucleation (Volmer–Weber mode). Such features are typical of Zone I in Thornton’s model, where low adatom mobility and geometric shadowing suppress lateral coalescence~\cite{Thornton,Thornton1}. The formation of this exceptionally fine amorphous nanocolumnar film is attributed to room-temperature deposition and RF magnetron sputtering, which lower adatom energy compared with DC sputtering and favor the development of narrow columns.

Lian et~al.~\cite{Lian} reported that Ti–Nb–Zr thin films deposited by DC magnetron sputtering at 150~$^{\circ}$C remained amorphous, though with wider columns. Similarly, Tallarico et~al.~\cite{Tal} observed column widths in the tens of nanometers for Ti–Nb–Zr films deposited at room temperature. In contrast, Gonzalez et~al.~\cite{Gon} found that increasing the substrate temperature to 200~$^{\circ}$C during DC/RF co-sputtering and applying a 30~V bias produced dense, crystalline films, while Liu et~al.~\cite{Liu} achieved nanocrystalline Ti–Nb–Zr coatings by increasing the adatom energy and lowering the argon pressure. These results collectively highlight that substrate temperature, bias, working pressure, and magnetron configuration strongly influence the column width and crystallinity of Ti–Nb–Zr-based thin films.

The columnar structure is advantageous for implant applications, as surface roughness increases the surface area and wettability, both of which are linked to the adsorption of proteins that promote the adhesion of osteogenic cells. However, not only the presence but also the width and morphology of columns are critical. The column width in our films ($\sim$15~nm), the smallest among the reviewed studies, falls within the range reported to enhance protein adsorption~\cite{Garcia}. For instance, Ercan et~al.~\cite{Ercan2013} observed that surface protrusions in the 5–19~nm range maximized fibronectin adsorption and also promoted the adsorption of type~IV collagen. Similarly, Scopelliti et~al.~\cite{Scop} reported that nanostructured silica surfaces with an average feature size of $\sim$15~nm exhibited the highest albumin and fibrinogen adsorption. The convex morphology of the columns may further help maintain the native conformation of adsorbed proteins, reducing the risk of denaturation~\cite{Xu}. In addition, the hierarchical structure of our films, combining nanocolumns with dome-like protrusions ($\sim$100~nm), provides multiscale topographical features relevant for future investigations of biointerface interactions, especially when combined with surface texturing in the hundreds-of-micrometers range to promote bone ingrowth.

SEM analysis of the as-deposited Ti–Nb–Zr–Ag film revealed a surface morphology similar to that of the Ag-free Ti–Nb–Zr film. A columnar structure was again observed in plan-view (Fig.~\ref{SEMasdep}c), with vertical columns and surface protrusions. The protrusions were less frequent and showed more irregular shapes, not always appearing hemispherical, but exhibited dimensions comparable to those observed in the Ag-free film. In cross-sectional view, shown in Fig.~\ref{SEMasdep}d, the general columnar architecture was retained, but in addition, round nanoparticles were visible throughout the film thickness. Angular surface particles were also present, resembling those observed in the Ag-free film.

Cross-sectional TEM confirmed the presence of vertically aligned columns extending through the entire film thickness (Fig.~\ref{TEMasdep}b). SAED likewise revealed a broad diffuse ring of lower intensity than in the Ag-free Ti–Nb–Zr film (Fig.~\ref{TEMasdep}b$^\prime$), suggesting a lower amorphous fraction in the Ag-containing film. In addition, discrete diffraction spots were observed, which were attributed partly to the silicon substrate and partly to crystalline phases within the film.

Two morphologically distinct crystalline features were present: column-integrated domains and particle-like crystalline domains embedded in an otherwise disordered matrix, suggesting multiple crystallization pathways during film growth. In Fig.~\ref{TEMasdep}b$^\prime$, the $\beta$-(Nb,Zr) phase is highlighted by green guide rings. Additional spots located in close proximity to these rings indicate the presence of other crystalline phases with similar interplanar spacings in the film. Consistent with the whole-thickness SAED indicating a $\beta$-(Nb,Zr) contribution, some of the column-integrated crystalline domains were identified as $\beta$-(Nb,Zr). A larger region, several tens of nanometres in size, near the silicon substrate and assigned to this phase in the $[001]$ orientation, is shown in Fig.~\ref{TEMasdep}c. The interplanar spacing $d_{110}$ was measured as $0.244~\mathrm{nm}$.

Another column-integrated phase was observed, whose measured $d$-spacings and interplanar angles are most consistent with the tetragonal AgZr ($P4/nmm$). Its key measured spacings are very close to those of \(\beta\)-(Nb,Zr): $d_{110}\big(\beta\text{-(Nb,Zr)}\big)=0.244~\mathrm{nm}$, $d_{110}(\mathrm{AgZr})=0.245~\mathrm{nm}$, and $d_{102}(\mathrm{AgZr})=0.238~\mathrm{nm}$, which explains the near-coincidence of the spot positions in Fig.~\ref{TEMasdep}b$^\prime$. Figure~\ref{TEMasdep}d shows a domain indexed as AgZr in the $[2\bar{2}\bar{1}]$ orientation. In the same field, a second phase appeared, showing a complete hexagonal set of $\{110\}$ reflections $60^{\circ}$ apart, consistent with a bcc lattice viewed along the $[\bar{1}11]$ zone axis. The measured spacing is $d_{110}=0.224~\mathrm{nm}$, corresponding to $a\approx0.317~\mathrm{nm}$, i.e., a lattice parameter $\sim$8~\% smaller than for $\beta$-(Nb,Zr) in this film. The bcc and AgZr domains were nearly co-oriented (offset by only a few degrees). This contracted-$a$ bcc domain was observed locally and does not appear in the whole-thickness SAED.

Figure~\ref{TEMasdep}e shows a nearly spherical Ag nanoparticle embedded in an otherwise amorphous matrix. In its vicinity, several few-nanometre crystalline seeds are present. Although these features share a common orientation (parallel lattice fringes), their measured spacings differ slightly from seed to seed, suggesting local variations of the lattice parameter due to compositional heterogeneity and/or elastic strain. According to the Fourier transform, these seeds are close to the $\beta$-(Nb,Zr) phase. In particular, slightly larger $d$-values (e.g., $d\approx0.248~\mathrm{nm}$) are consistent with local Zr enrichment.

This structure may strongly influence the mechanical properties of the coating, as nanoscale grains can act as barriers to local shear-band propagation in the amorphous matrix and to dislocation motion within the crystalline domains. Amorphous-nanocrystalline composites have been reported to exhibit significantly higher hardness and wear resistance compared to fully amorphous or crystalline structures~\cite{Xia,Liu2} due to the Hall–Petch effect~\cite{HallPetch}, interface strengthening, and, in addition, local variations in cohesive energy arising from the high degree of local chemical disorder in MEAs.~\cite{Feltrin}.

\subsection {Chemical characterization}
Despite the equimolar composition of the Ti–Nb–Zr sputtering target, compositional analysis by STEM EDX revealed a significant deviation from the nominal stoichiometry in the resulting film. The Ti–Nb–Zr film exhibited a reduced Ti content (26~at.\%), accompanied by elevated levels of Nb (36~at.\%) and Zr (38 at.\%). In the Ti–Nb–Zr–Ag thin film, EDX analysis showed an elemental distribution of 13~at.\% Ag, 16~at.\% Ti, 35~at.\% Zr, and 36~at.\% Nb, further emphasizing the preferential incorporation of Zr and Nb at the expense of Ti.
This discrepancy may be attributed to differences in sputtering yields and surface binding energies among the constituent elements. Ti can exhibit a sputtering yield lower than that of Nb and Zr because of its lower atomic mass, which reduces the efficiency of momentum transfer.

\begin{figure*}[t]
\centering
\includegraphics[width=1.03\textwidth]{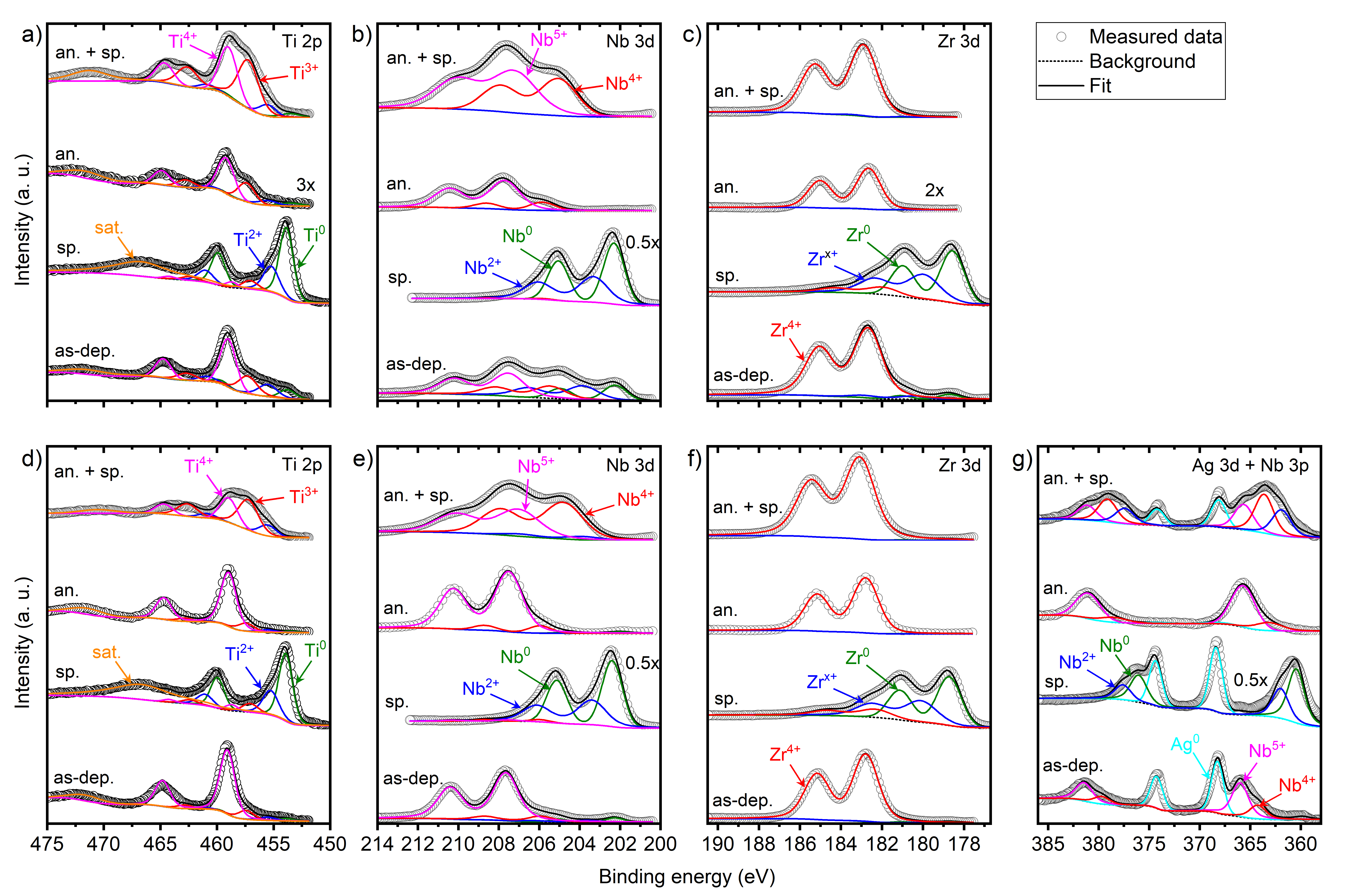}
\caption{\label{XPS} XPS spectra of Ti $2p$, Nb $3d$, and Zr $3d$ core levels for Ti--Nb--Zr (a–c) and Ti $2p$, Nb $3d$, Zr $3d$, and Ag $3d$ + Nb $3p$ for Ti–Nb–Zr–Ag (d--g). Spectra are shown for as-deposited, sputtered, annealed, and annealed + sputtered states (bottom to top).}
\end{figure*}

 XPS analysis revealed distinct changes in surface chemistry across different sample states. In the as-deposited condition, the surfaces of both alloys were strongly oxidized. Figure~\ref{XPS}a shows Ti $2p$ spectra with the dominant Ti$^{4+}$ peak at position 459.01~eV, accompanied by weaker contributions from Ti$^{3+}$, Ti$^{2+}$, and Ti$^{0}$ at 457.29 eV, 455.50~eV, and 453.77 eV, respectively. These peaks correspond to TiO$_2$, Ti$_2$O$_3$, and TiO oxide phases, and metallic Ti~\cite{Biesinger2010, Lopez2011}. After Ar$^+$ sputtering, oxide contributions were significantly reduced and metallic Ti became dominant, although residual oxide signals remained. 
 
 For the annealed sample, the surface remained oxidized even after sputtering, with a slightly enhanced signal from Ti$^{3+}$. Fig.~\ref{XPS}b presents the Nb $3d$ spectra, where four nearly equally intense components were observed in the as-deposited state: Nb$^{5+}$ (207.52~eV), Nb$^{4+}$ (205.45~eV), Nb$^{2+}$ (203.85~eV), and Nb$^{0}$ (202.24~eV), corresponding to Nb$_2$O$_5$, NbO$_2$, NbO, and metallic Nb~\cite{Moulder1995XPSHandbook, Islam2022, Lopez2011}. Following sputtering, only Nb$^{2+}$ and Nb$^{0}$ peaks were detected. The sputtering of the annealed sample mirrored the Ti $2p$ behavior, yielding Nb$^{5+}$ and Nb$^{4+}$ states.
In contrast, the Zr $3d$ spectrum of the as-deposited sample (Fig.~\ref{XPS}c) was dominated by Zr$^{4+}$ at 182.68~eV (ZrO$_2$), while substoichiometric ZrO$_x$ and metallic Zr$^{0}$ components at 179.97~eV and 178.56~eV became evident only after sputtering~\cite{Chua2003, Liao2010}. However, the annealed sample showed only the Zr$^{4+}$ state, which remained unchanged after sputtering, suggesting that Zr had oxidized deeper into the bulk compared to the other elements. In addition to the oxidation states mentioned, more complex oxides, such as those described by Lee and Ryu~\cite{Lee2017}, cannot be excluded, as they occur at similar binding energies.
The extensive oxidation observed after annealing was presumably promoted by native surface oxides present on the as-deposited films and by residual oxygen in the vacuum chamber during heat treatment.

The behavior of the Ti--Nb--Zr--Ag sample (Fig.~\ref{XPS}e–g) was similar; however, the presence of Ag resulted in less pronounced oxidation of Ti and Zr after annealing and sputtering, indicating an improved oxidation resistance of the alloy. The Ag $3d$ peak at 368.3~eV~\cite{Moulder1995XPSHandbook} consistently indicated metallic Ag and remained unchanged before and after sputtering, confirming that Ag did not oxidize. Notably, this peak disappeared in the annealed state, suggesting the absence of Ag at the surface and its presence at deeper regions of the sample.

\subsection {Microstructure and morphology of annealed films}

\begin{figure*}[t]
\centering
\includegraphics[width=0.325\textwidth]{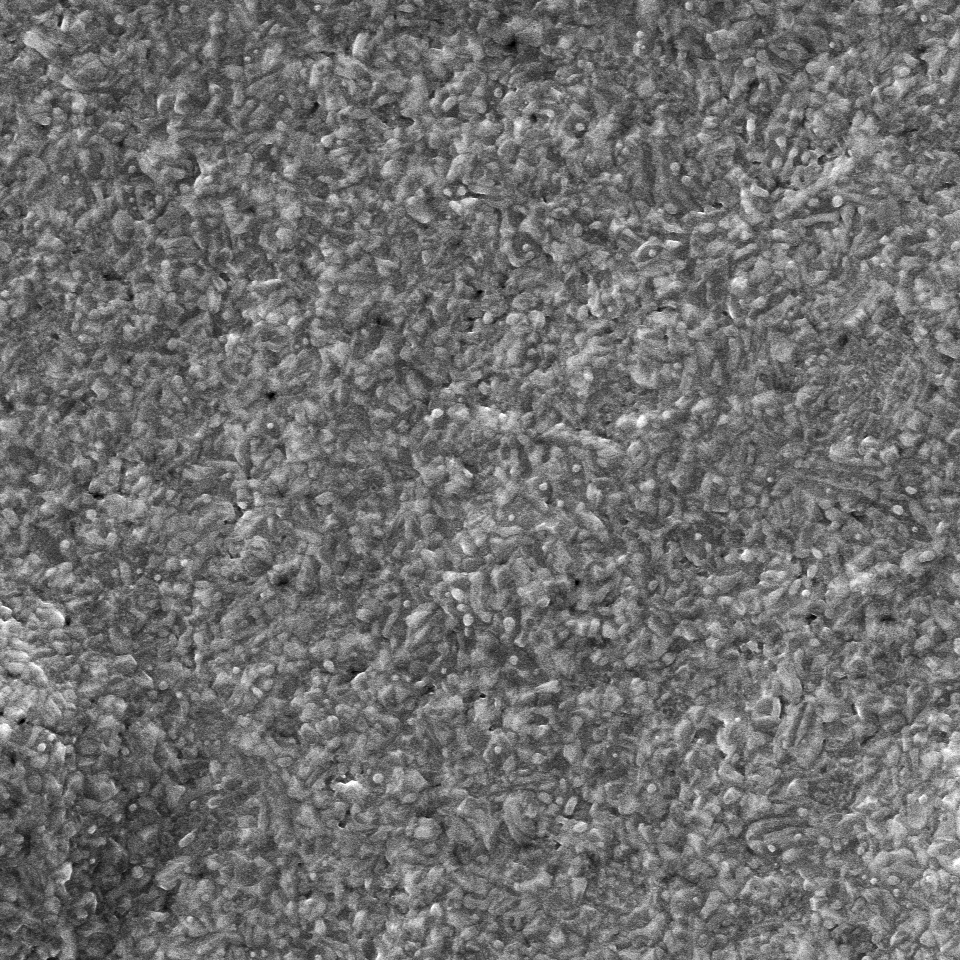}\put(-0.3\textwidth,0.02\textwidth){\color{white}\rule{0.0764\textwidth}{0.0035\textwidth}}\hspace{0.0125\textwidth}%
\put(-0.3\textwidth,0.03\textwidth){\sf\color{white}2 $\upmu$m}\hspace{0.0125\textwidth}%
\put(-0.315\textwidth,0.296\textwidth){\color{white}{\normalsize\textbf a)}}\hspace{0.0125\textwidth}%
\includegraphics[width=0.325\textwidth]{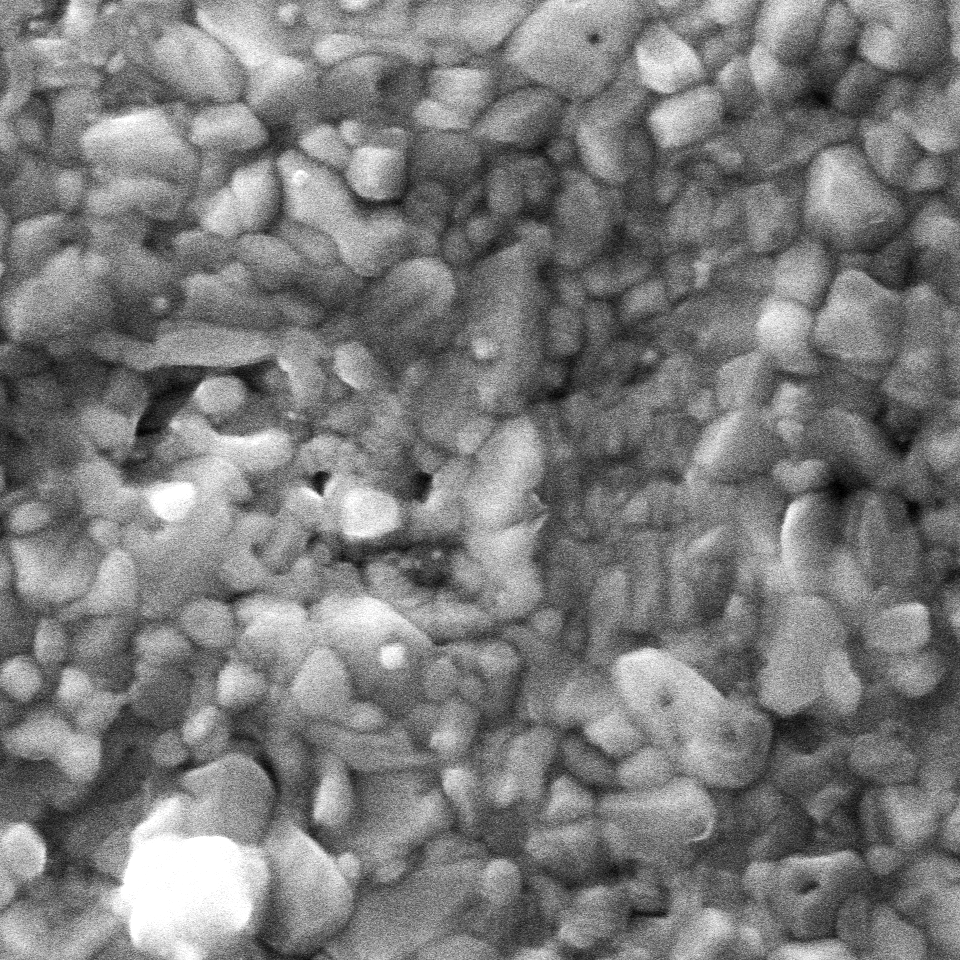}\put(-0.3\textwidth,0.02\textwidth){\color{black}\rule{0.0865\textwidth}{0.0035\textwidth}}\hspace{0.0125\textwidth}%
\put(-0.3\textwidth,0.03\textwidth){\sf\color{black}500 nm}\hspace{0.0125\textwidth}%
\put(-0.315\textwidth,0.296\textwidth){\color{white}{\normalsize\textbf b)}}\hspace{0.0125\textwidth}%
\includegraphics[width=0.325\textwidth]{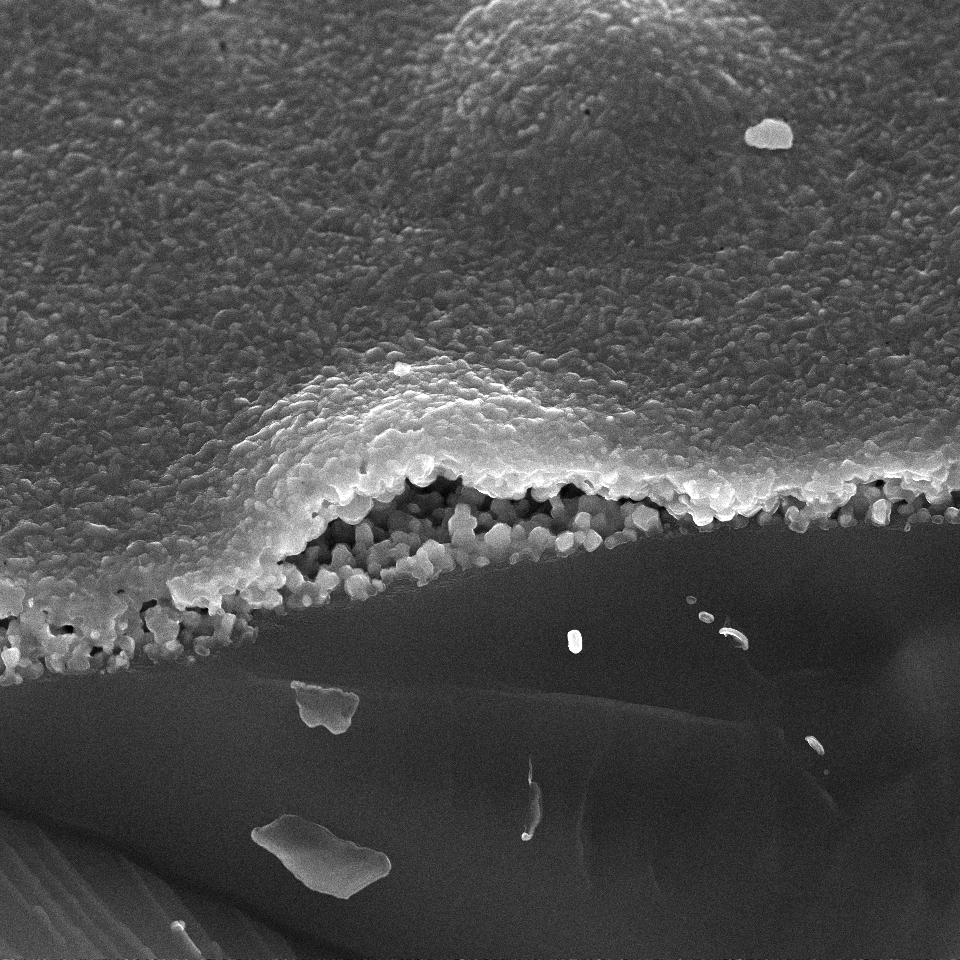}\put(-0.3\textwidth,0.02\textwidth){\color{white}\rule{0.074\textwidth}{0.0035\textwidth}}
\vspace{0.01\textwidth}%
\put(-0.3\textwidth,0.03\textwidth){\sf\color{white}2 $\upmu$m}%
\put(-0.315\textwidth,0.296\textwidth){\color{white}{\normalsize\textbf c)}}%

\includegraphics[width=0.325\textwidth]{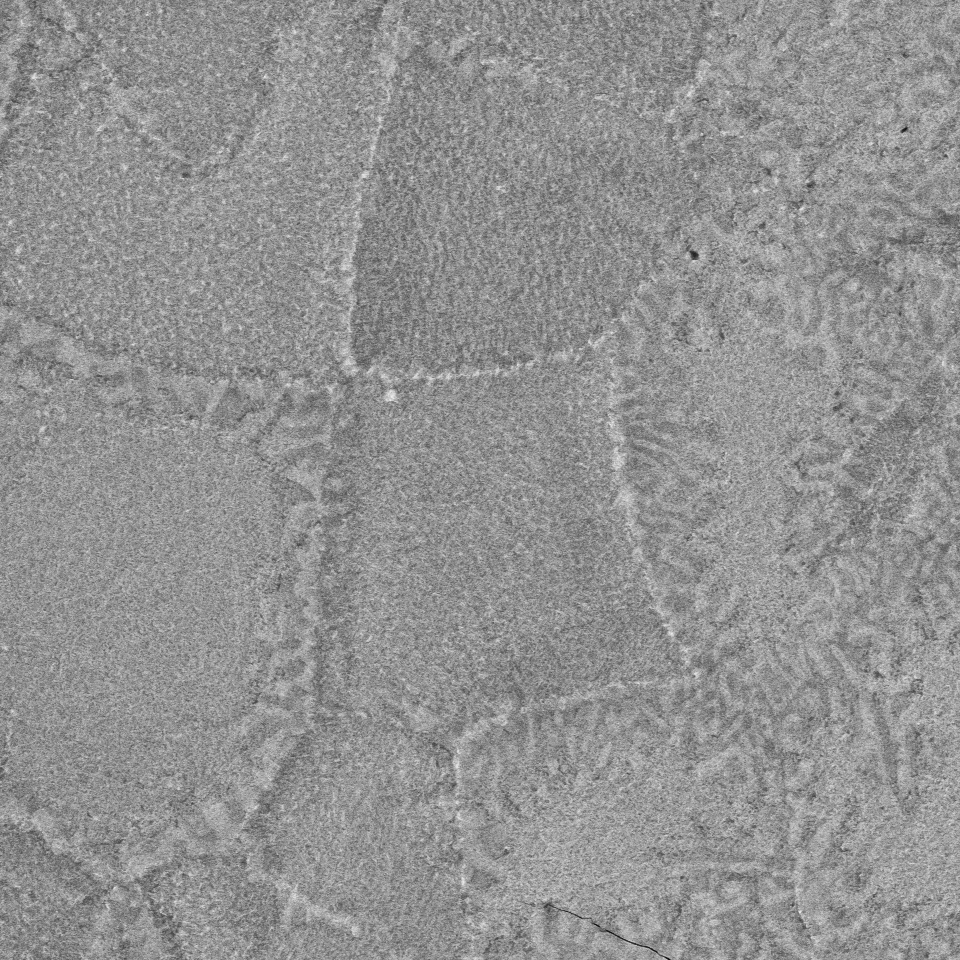}\put(-0.3\textwidth,0.02\textwidth){\color{white}\rule{0.0764\textwidth}{0.0035\textwidth}}\hspace{0.0125\textwidth}
\put(-0.3\textwidth,0.03\textwidth){\sf\color{white}2 $\upmu$m}\hspace{0.0125\textwidth}%
\put(-0.315\textwidth,0.296\textwidth){\color{white}{\normalsize\textbf d)}}\hspace{0.0125\textwidth}%
\includegraphics[width=0.325\textwidth]{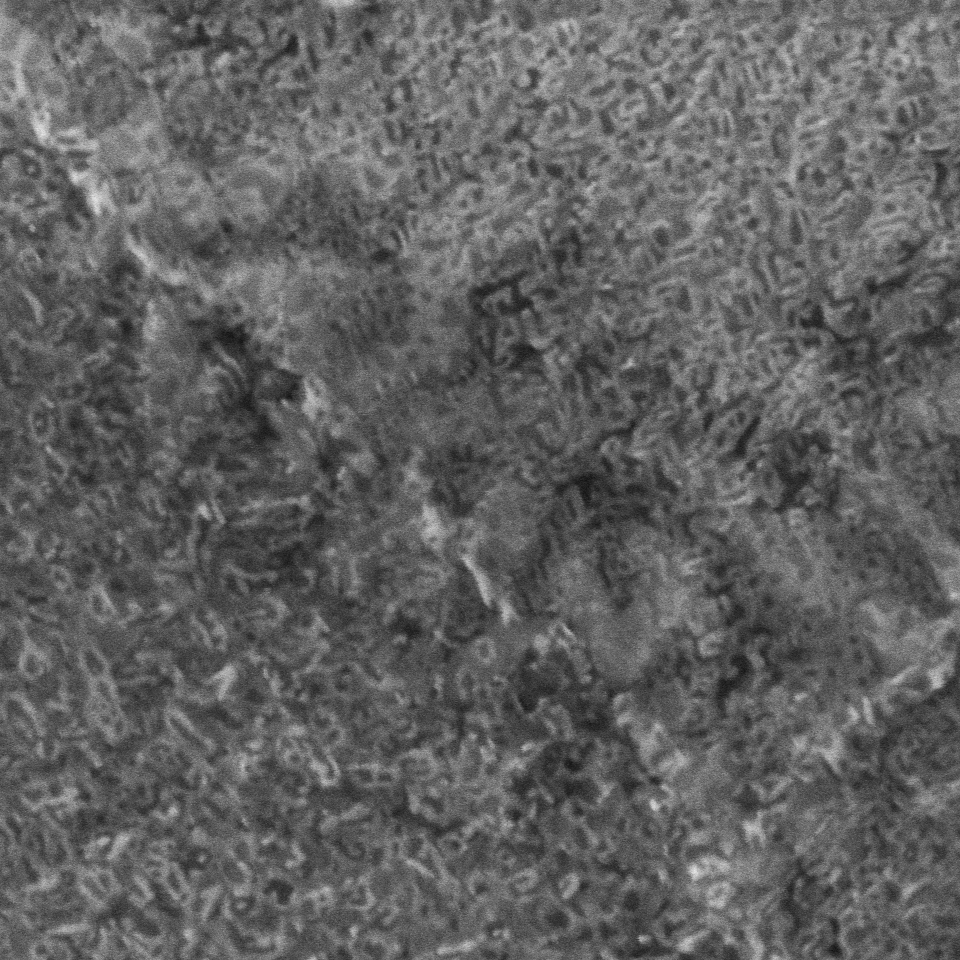}\put(-0.3\textwidth,0.02\textwidth){\color{white}\rule{0.0865\textwidth}{0.0035\textwidth}}\hspace{0.0125\textwidth}%
\put(-0.3\textwidth,0.03\textwidth){\sf\color{white}500 nm}\hspace{0.0125\textwidth}%
\put(-0.315\textwidth,0.296\textwidth){\color{white}{\normalsize\textbf e)}}\hspace{0.0125\textwidth}%
\includegraphics[width=0.325\textwidth]{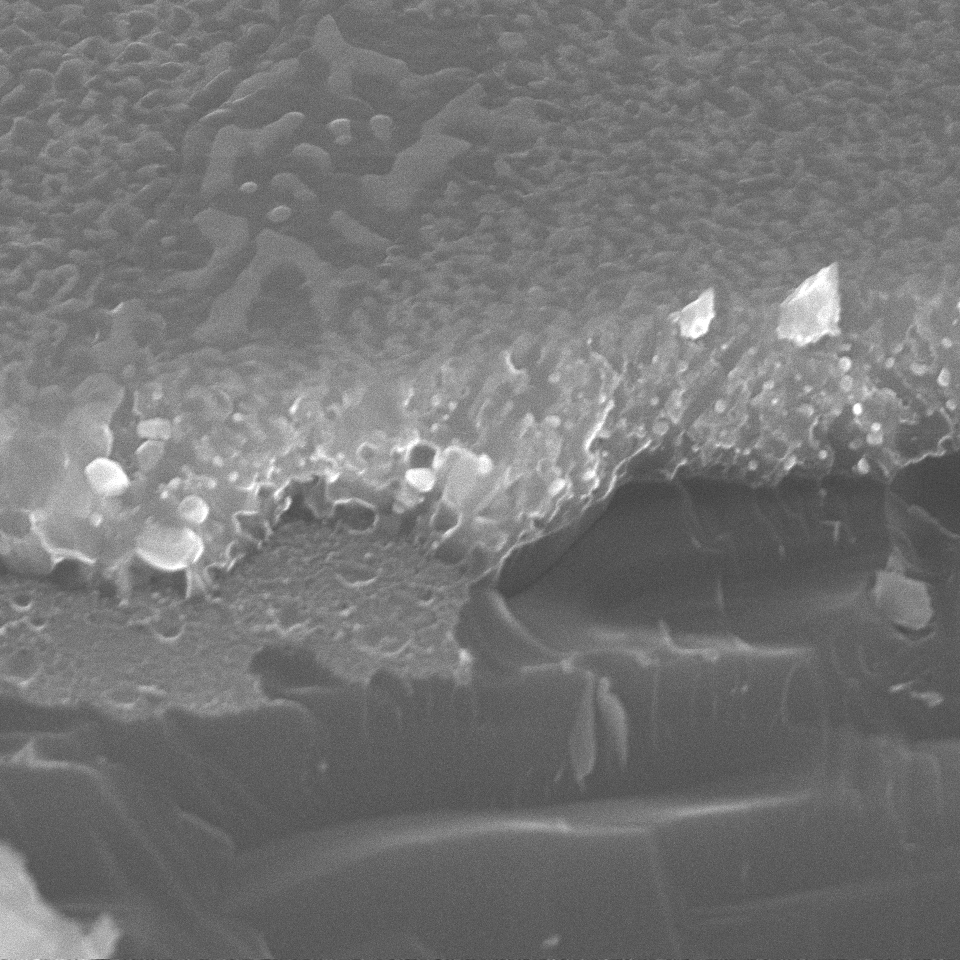}\put(-0.3\textwidth,0.02\textwidth){\color{white}\rule{0.06\textwidth}{0.0035\textwidth}}
\put(-0.3\textwidth,0.03\textwidth){\sf\color{white}500 nm}%
\put(-0.315\textwidth,0.292\textwidth){\color{white}{\normalsize\textbf f)}}%
\caption{\label{annealed} \small  SEM micrographs of Ti–Nb–Zr (a–c) and Ti–Nb–Zr–Ag (d–f) thin films after annealing. (a,d) plan-view, lower magnification; (b,e) plan-view, higher magnification; (c,f) cross-section, 45$^{\circ}$ tilt.} 
\end{figure*}

SEM analysis revealed that, although the as-deposited Ti–Nb–Zr and Ti–Nb–Zr–Ag thin films exhibited a very similar columnar morphology, this feature was no longer preserved after annealing, and the subsequent microstructural evolution differed significantly between the two films.

\begin{figure*}[t]
\centering
\includegraphics[width=0.24\textwidth]{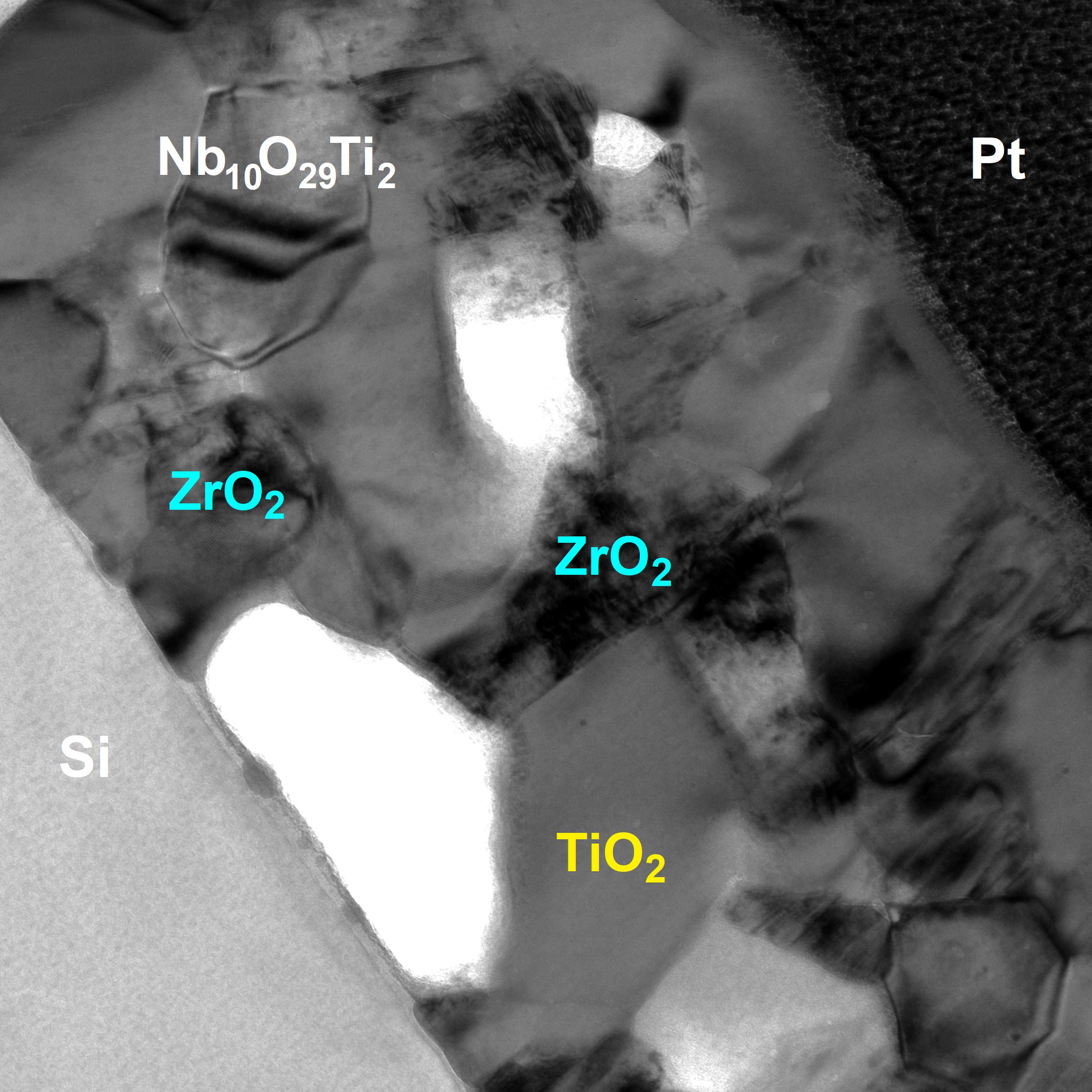}\put(-0.23\textwidth,0.01\textwidth){\color{white}\rule{0.0598\textwidth}{0.0035\textwidth}}\hspace{0.0125\textwidth}%
\put(-0.23\textwidth,0.02\textwidth){\small\sf\color{white}200 nm}\hspace{0.0125\textwidth}%
\put(-0.234\textwidth,0.216\textwidth){\color{white}{\normalsize\textbf a)}}\hspace{0.005\textwidth}%
\includegraphics[width=0.24\textwidth]{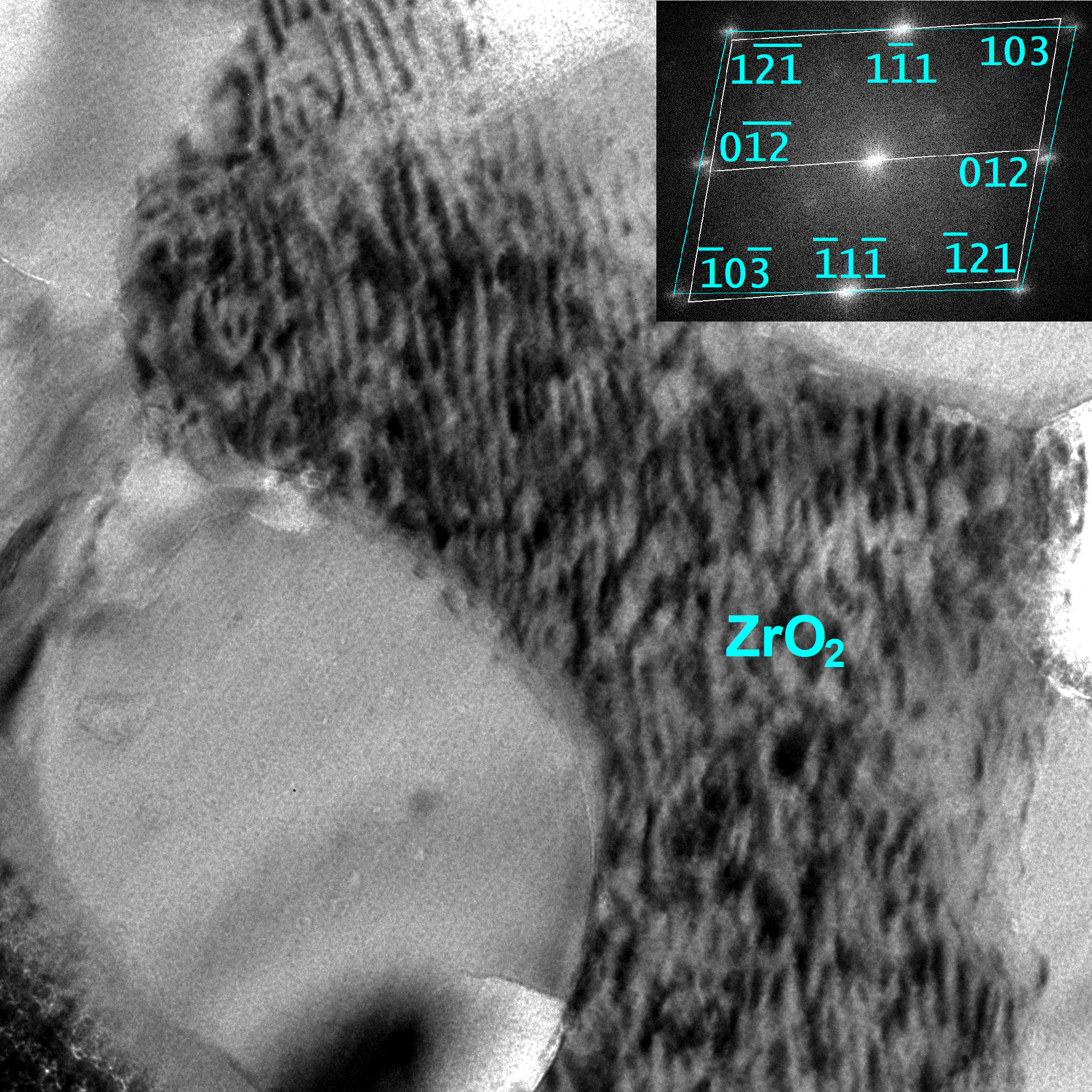}\put(-0.23\textwidth,0.01\textwidth){\color{white}\rule{0.065\textwidth}{0.0035\textwidth}}\hspace{0.0125\textwidth}%
\put(-0.23\textwidth,0.02\textwidth){\small\sf\color{white}100 nm}\hspace{0.0125\textwidth}%
\put(-0.234\textwidth,0.216\textwidth){\color{white}{\normalsize\textbf b)}}\hspace{0.0125\textwidth}%
\includegraphics[width=0.24\textwidth]{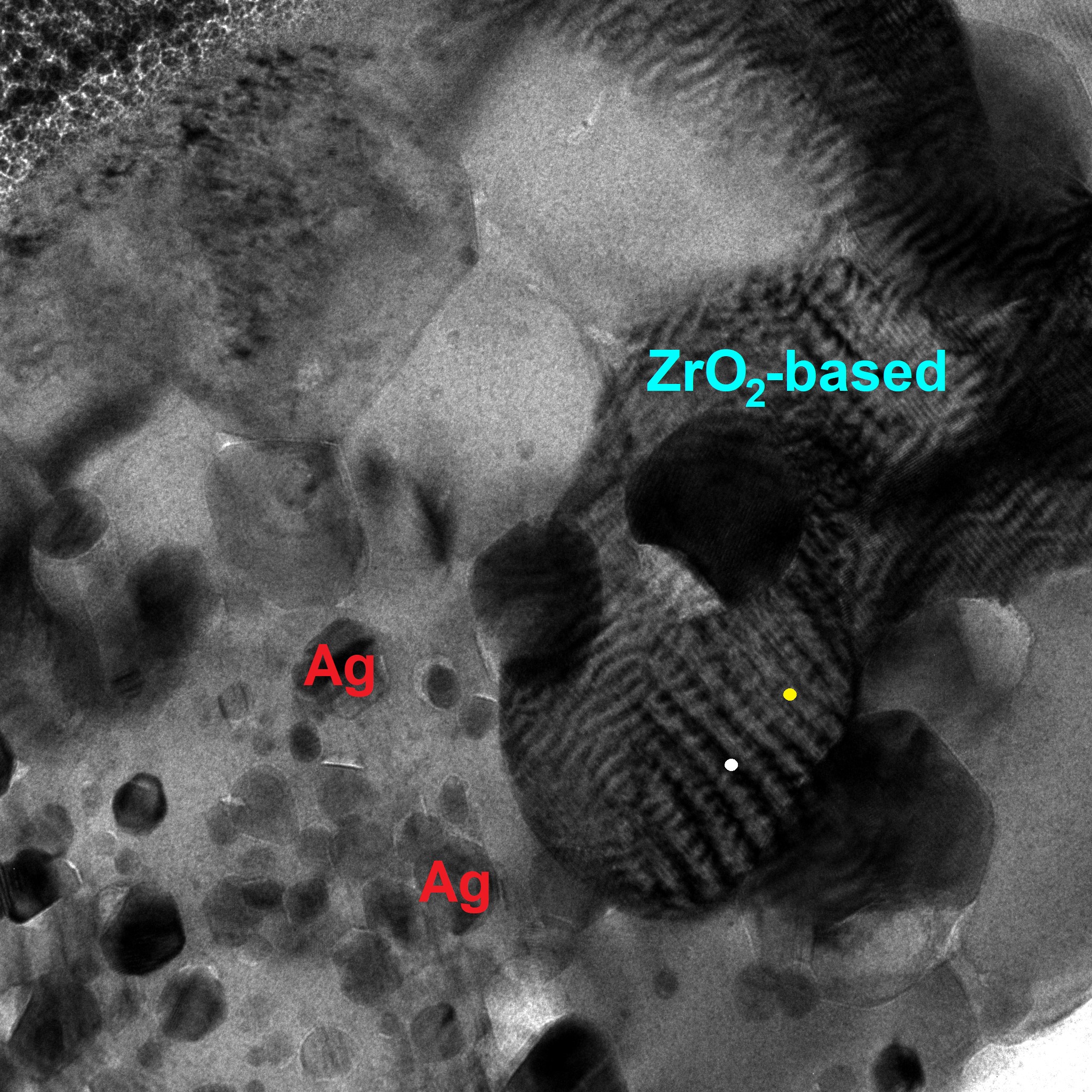}\put(-0.23\textwidth,0.01\textwidth){\color{white}\rule{0.0634\textwidth}{0.0035\textwidth}}
\put(-0.23\textwidth,0.02\textwidth){\small\sf\color{white}100 nm}%
\put(-0.234\textwidth,0.216\textwidth){\color{white}{\normalsize\textbf e)}}\hspace{0.005\textwidth}%
\includegraphics[width=0.24\textwidth]{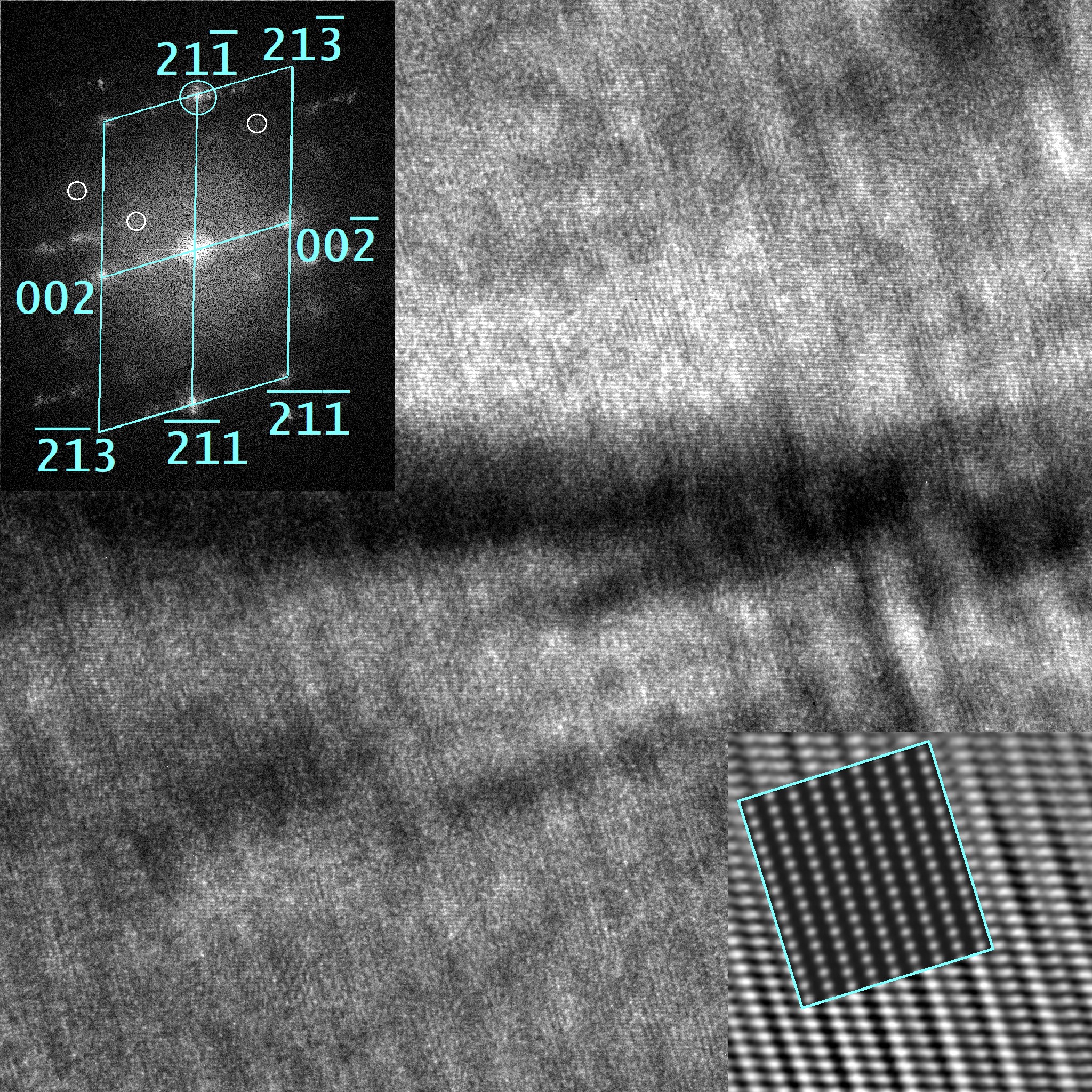}\put(-0.23\textwidth,0.01\textwidth){\color{white}\rule{0.061\textwidth}{0.0035\textwidth}}\hspace{0.0125\textwidth}%
\put(-0.23\textwidth,0.02\textwidth){\small\sf\color{white}10 nm}
\put(-0.234\textwidth,0.216\textwidth){\color{white}{\normalsize\textbf f)}}%
\vspace{0.002\textwidth}%

\includegraphics[width=0.24\textwidth]{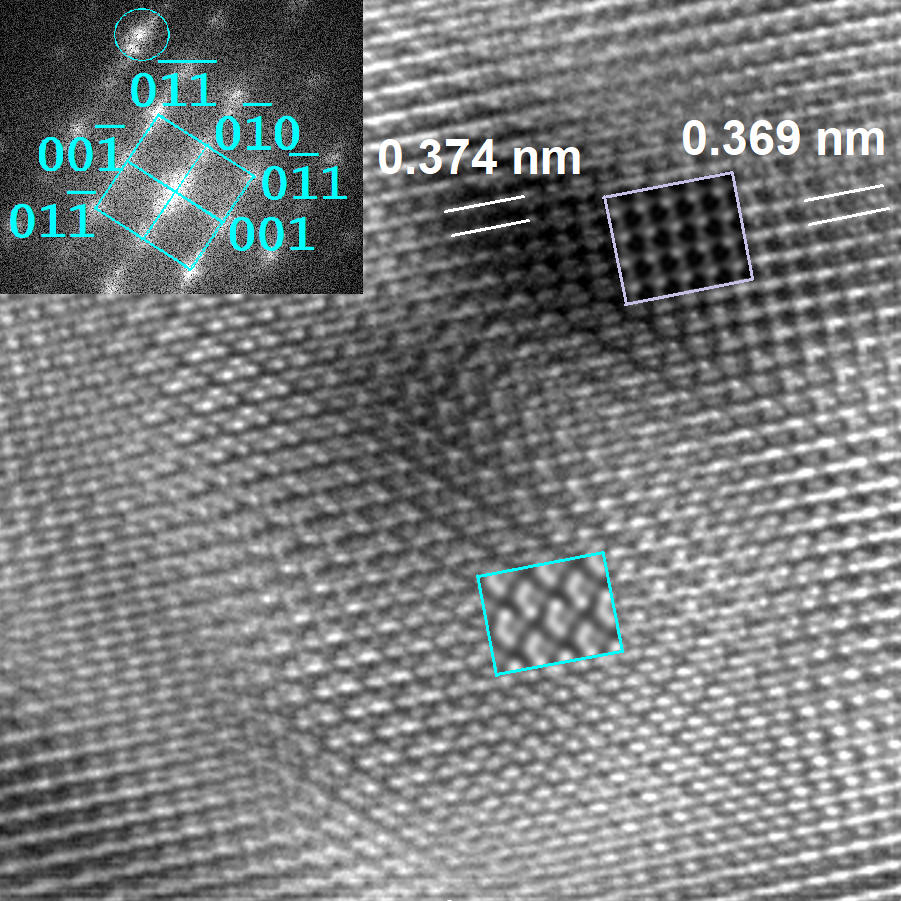}\put(-0.23\textwidth,0.01\textwidth){\color{white}\rule{0.072\textwidth}{0.0035\textwidth}}\hspace{0.0125\textwidth}%
\put(-0.23\textwidth,0.02\textwidth){\small\sf\color{white}5 nm}\hspace{0.0125\textwidth}%
\put(-0.234\textwidth,0.216\textwidth){\color{white}{\normalsize\textbf c)}}\hspace{0.005\textwidth}%
\includegraphics[width=0.24\textwidth]{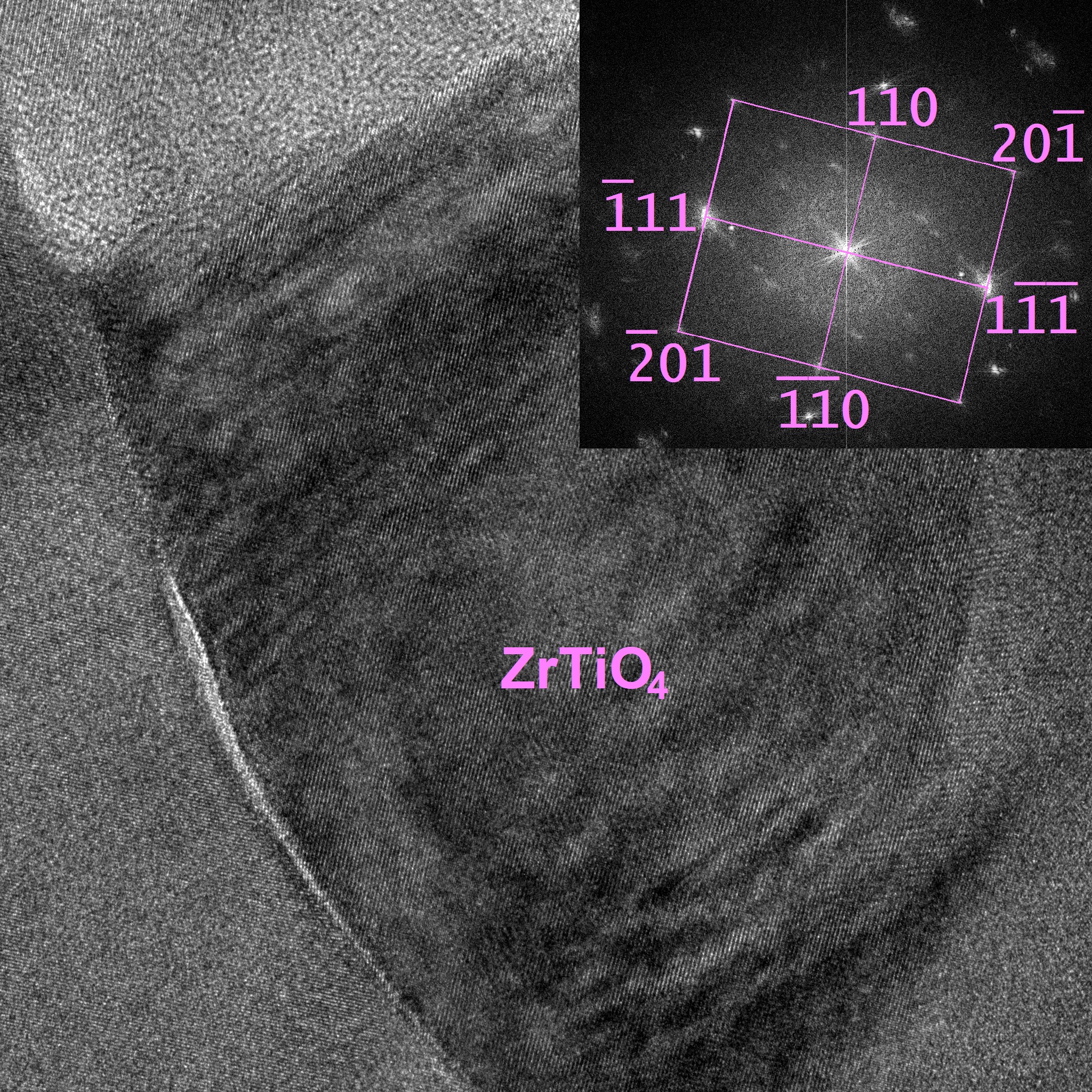}\put(-0.23\textwidth,0.01\textwidth){\color{white}\rule{0.0574\textwidth}{0.0035\textwidth}}\hspace{0.0125\textwidth}%
\put(-0.23\textwidth,0.02\textwidth){\small\sf\color{white}20 nm}\hspace{0.0125\textwidth}%
\put(-0.234\textwidth,0.216\textwidth){\color{white}{\normalsize\textbf d)}}\hspace{0.0125\textwidth}%
\includegraphics[width=0.24\textwidth]{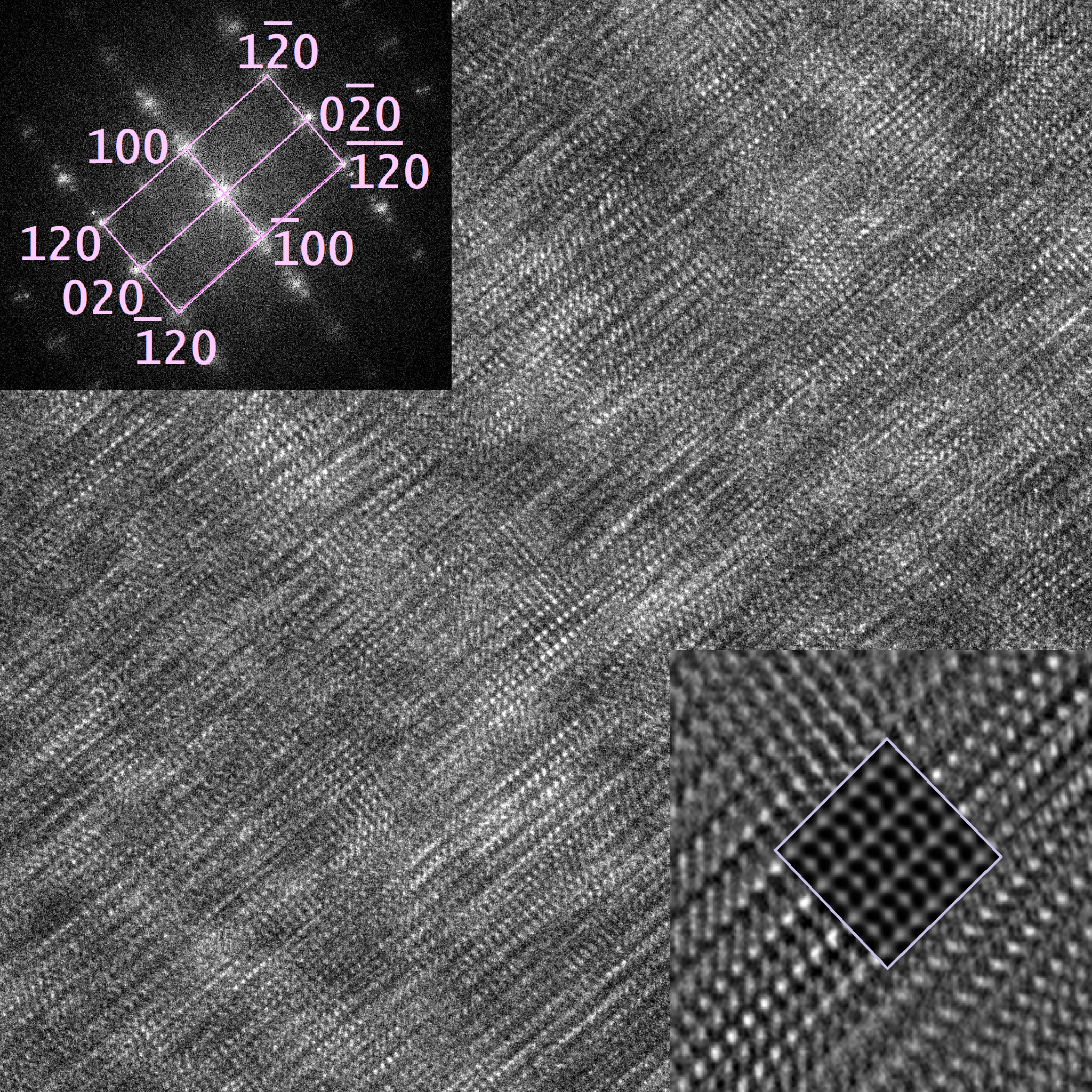}\put(-0.23\textwidth,0.01\textwidth){\color{white}\rule{0.0774\textwidth}{0.0036\textwidth}}
\vspace{0.01\textwidth}%
\put(-0.23\textwidth,0.02\textwidth){\sf\color{white}10 nm}%
\put(-0.234\textwidth,0.216\textwidth){\color{white}{\normalsize\textbf g)}}\hspace{0.005\textwidth}%
\includegraphics[width=0.24\textwidth]{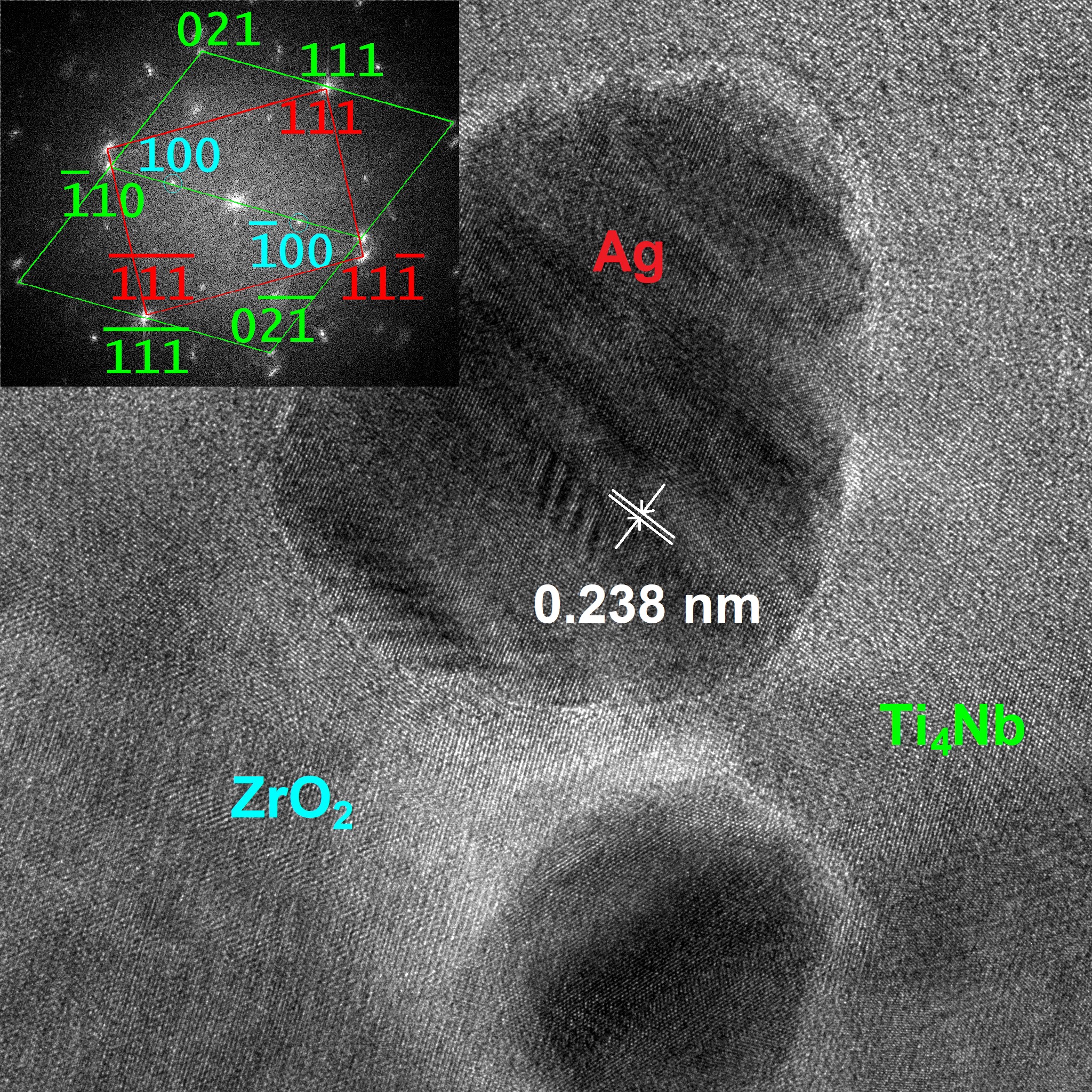}\put(-0.23\textwidth,0.01\textwidth){\color{white}\rule{0.0736\textwidth}{0.0035\textwidth}}
\vspace{0.005\textwidth}%
\put(-0.23\textwidth,0.02\textwidth){\small\sf\color{white}20 nm}%
\put(-0.234\textwidth,0.216\textwidth){\color{white}{\normalsize\textbf h)}}\vspace{-0.0125\textwidth}%

\includegraphics[width=0.118\textwidth]{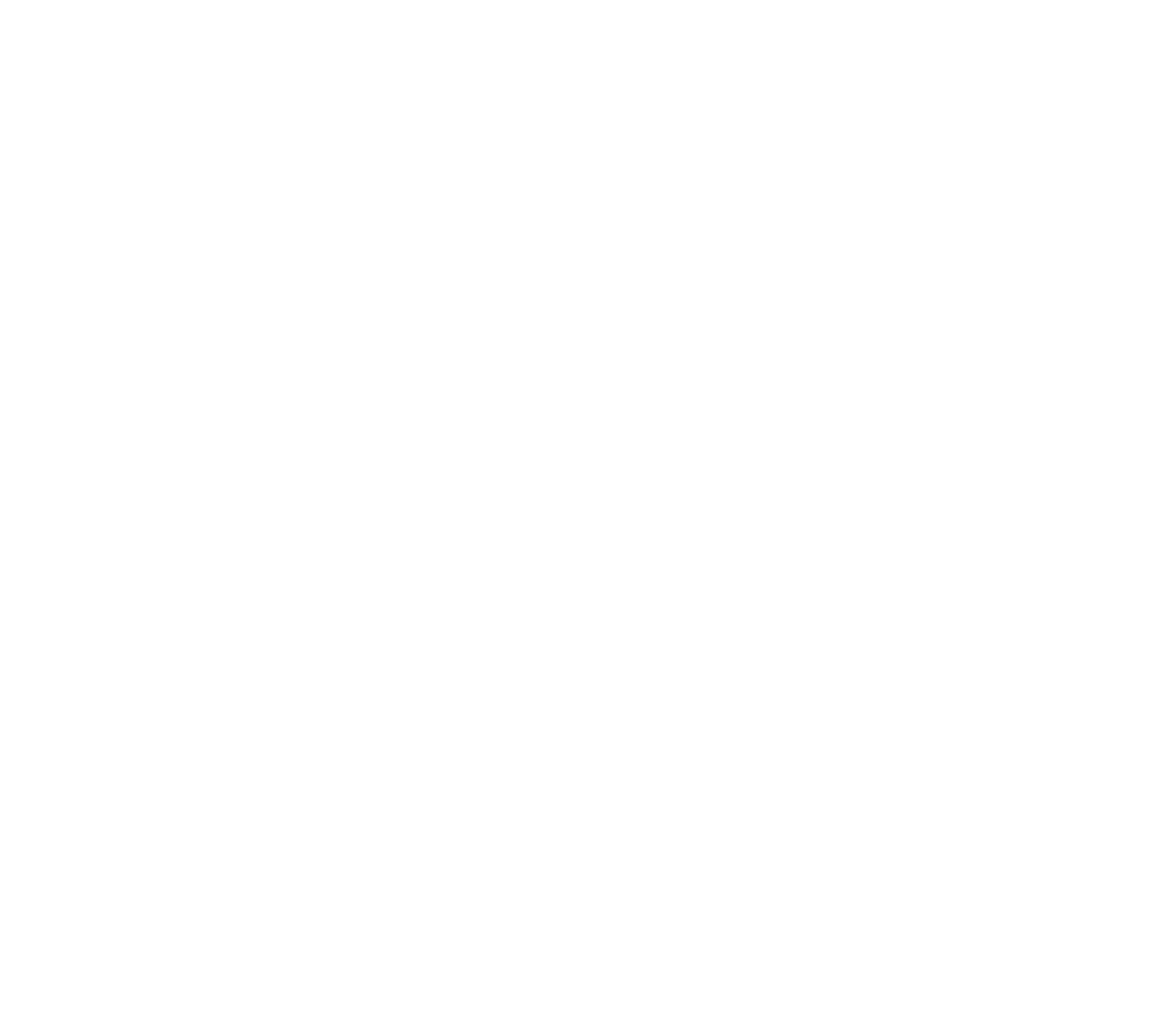}\hspace{0.005\textwidth}%
\includegraphics[width=0.118\textwidth]{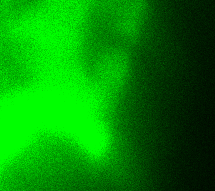}\put(-0.112\textwidth,0.08\textwidth){\color{white}{\footnotesize Ti}}\hspace{0.005\textwidth}%
\includegraphics[width=0.118\textwidth]{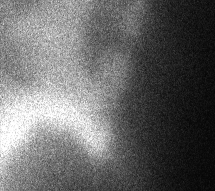}\put(-0.112\textwidth,0.08\textwidth){\color{white}{\footnotesize Nb}}\hspace{0.005\textwidth}%
\includegraphics[width=0.118\textwidth]{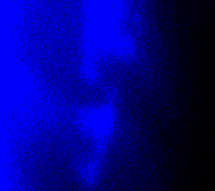}\put(-0.112\textwidth,0.08\textwidth){\color{white}{\footnotesize Zr}}\hspace{0.011\textwidth}%
\includegraphics[width=0.118\textwidth]{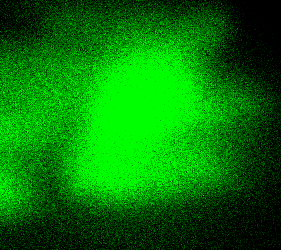}\put(-0.112\textwidth,0.08\textwidth){\color{white}{\footnotesize Ti}}\hspace{0.005\textwidth}%
\includegraphics[width=0.118\textwidth]{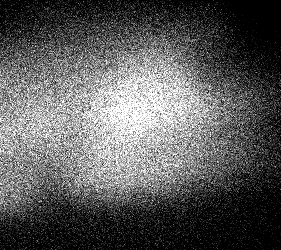}\put(-0.112\textwidth,0.08\textwidth){\color{white}{\footnotesize Nb}}\hspace{0.005\textwidth}%
\includegraphics[width=0.118\textwidth]{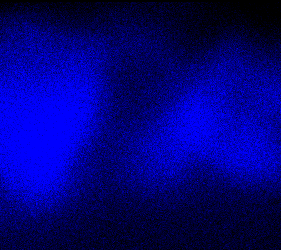}\put(-0.112\textwidth,0.08\textwidth){\color{white}{\footnotesize Zr}}\hspace{0.005\textwidth}%
\includegraphics[width=0.118\textwidth]{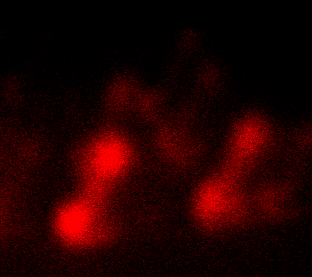}\put(-0.112\textwidth,0.08\textwidth){\color{white}{\footnotesize Ag}}%
\caption{\label{TEMannealed} \small 
Cross-sectional TEM of annealed Ti–Nb–Zr (left) and Ti–Nb–Zr–Ag (right) thin films. 
(a) Overview with labelled phases Nb$_{10}$O$_{29}$Ti$_2$, ZrO$_2$, and TiO$_2$; 
(b) Monoclinic ZrO$_2$ [$3 2\bar{1}$] with FFT inset; 
(c) Inverse FFT from monoclinic ZrO$_2$ [$100$] and simulated HRTEM images: monoclinic ZrO$_2$ (turquoise frame) and orthorhombic variant (pink frame), differences in $d$-spacings highlighted. Inset: FFT with circled satellite reflections; 
(d) ZrTiO$_4$ [$1 \bar{1} 2$] with FFT inset; 
(e) Overview with labelled Ag and ZrO$_2$-based phases; 
(f) Tetragonal ZrO$_2$-based [$1 \bar{2} 0$]; FFT (upper inset) with circled diffuse reflections of early-stage monoclinic/orthorhombic ZrO$_2$; inverse FFT with simulated image in turquoise frame (lower inset);
(g) Orthorhombic ZrO$_2$ $[001]$; FFT (upper inset), inverse FFT with simulated image in pink frame (lower inset); 
(h) Ag [$1 \bar{1} 0$], Ti$_4$Nb [$\bar{1}\bar{1} 2$], and monoclinic ZrO$_2$ with FFT inset. 
Bottom: elemental maps, with map widths of approximately 700~nm for Ti–Nb–Zr and 600~nm for Ti–Nb–Zr–Ag.}
\end{figure*}

In the Ti–Nb–Zr film, both cross-sectional and plan-view SEM observations revealed a granular, highly porous structure with loosely packed, angular grains up to $\sim$400~nm in size, and an increased surface roughness. Moreover, bubble-like intralayer cavities with parts of the layer remaining attached to the substrate were found, indicating intralayer decohesion (Fig.~\ref{annealed}a–c). TEM cross-sectional imaging (Fig.~\ref{TEMannealed}a) confirmed the porous granular morphology, with some regions exhibiting missing grains (pull-outs), most likely caused by grain-boundary separation. The SAED patterns revealed a fully crystalline structure containing several phases, most of which were identified as Zr-, Nb-, and Ti-based oxides. In the same panel, representative grains are labelled, revealing the presence of monoclinic baddeleyite ZrO$_2$ ($P2/c$), monoclinic Nb$_{10}$O$_{29}$Ti$_2$ ($A2/m$), and tetragonal rutile TiO$_2$ ($P4_2/mnm$). In rare cases, an orthorhombic Ti$_4$Nb phase ($Cmcm$) was also detected. 

Some of the ZrO$_2$ grains exhibited an interesting mottled, tweed-like contrast with elongated dark features lying parallel to the (012) planes (Fig.~\ref{TEMannealed}b). The corresponding Fourier transform revealed two slightly rotated reciprocal lattices, with one set of reflections matching monoclinic ZrO$_2$ ($a = 0.525$~nm, $b = 0.530$~nm, $c = 0.540$~nm, $\beta \approx 93^{\circ}$) and the other corresponding to a residual orthorhombic ($Pca2_1$-like) variant with slightly contracted cell parameters ($a = 0.521$~nm, $b = 0.526$~nm, $c = 0.535$~nm, $\beta \approx 89^{\circ}$). EDX analyses revealed a Nb-substituted solid solution (Zr:Nb $\approx$ 2:1) with local compositional fluctuations and oxygen deficiency ($\sim$50~at.\% O), which likely caused the observed contrast modulation. In some regions (Fig.~\ref{TEMannealed}c), periodic variations in the $d$-spacing of the (011) planes ($d_{011}$ = 0.369–0.374~nm) were accompanied by satellite reflections symmetrically positioned on both sides of the main spots in the FFT, indicating nanoscale compositional and structural modulation within the oxide lattice, where orthorhombic $Pca2_1$ domains remained embedded in the monoclinic ZrO$_2$ matrix.

Elemental EDX mapping acquired at lower magnification revealed a general segregation trend induced by annealing: Ti and Nb tended to cluster together, whereas Zr segregated separately. Nevertheless, areas with partial intermixing of the elements were also observed, and e.g. a mixed ortorhombic oxide ZrTiO$_4$ ($Pbcn$) was identified in such a region (Fig.~\ref{TEMannealed}d).

In plan-view SEM of the Ti–Nb–Zr–Ag thin film, domains several micrometers in size were observed, separated by distinct boundaries and exhibiting a fine striped contrast. The stripe orientation varied between domains, suggesting that the contrast originates from nanoscale chemical modulation combined with crystallographic orientation differences. Compared with the Ag-free film, the surface appeared smoother and the film remained compact, with no significant grain separation or intralayer cavities; only occasional cracks were present, typically located at domain boundaries.

Figure~\ref{TEMannealed}e shows a representative cross-sectional microstructure of the Ti–Nb–Zr–Ag thin film. In contrast to the Ag-free film, where grains were of comparable size, the grain size here varied significantly, ranging from small, nearly equiaxed Ag nanoparticles (tens of nanometers) to large regions (hundreds of nanometers) exhibiting a characteristic striped contrast similar to the Ag-free film. A point analysis performed along a line perpendicular to the wide stripes revealed that all metallic elements and oxygen were present within this phase. The darker stripes were enriched in Zr and Ag, whereas the brighter regions contained higher concentrations of Nb and Ti. Specifically, at the white point the metallic composition was 18.8~at.\% Ti, 33.7~at.\% Zr, 37.1~at.\% Nb, and 10.5~at.\% Ag, while at the yellow point it was 26.4~at.\% Ti, 23.5~at.\% Zr, 43.5~at.\% Nb, and 6.6~at.\% Ag. The overall oxygen content was approximately 50~at.\%. This structure was identified as a tetragonal ZrO$_2$-based solid solution ($P4_2/nmc$) oriented close to [1$\bar{2}$0], in which atoms of other metallic elements are likely present substitutionally, causing a slight contraction of the lattice parameters ($a \approx 0.35$~nm, $c \approx 0.49$~nm) compared with database values. The presence of these substitutional elements, together with a considerable oxygen deficiency, is considered to be responsible for the stabilization of this metastable tetragonal phase. In contrast, individual grains of monoclinic ZrO$_2$ were also identified in the film, characterized by a strong dominance of Zr.

Figure~\ref{TEMannealed}f shows a detailed view revealing not only the wide stripes (the horizontal band in the middle of the image) lying parallel to the (2$\bar{1}$1) planes, but also the finer modulations oriented along the (001) planes. The positions of the very broad and diffuse reflections in the corresponding FFT pattern suggest an early-stage structural organization toward monoclinic or orthorhombic ZrO$_2$, although the pronounced broadening prevents unambiguous phase assignment. In some regions, the (100) planes appeared wavy (Fig.~\ref{TEMannealed}g), accompanied by strong satellite reflections in the corresponding FFT pattern. These features indicate local chemical modulations that lead to slight changes in the lattice parameter within the oxide lattice. The structure in this region was identified as orthorhombic ZrO$_2$ ($Pca2_1$). Figure~\ref{TEMannealed}h shows an Ag nanoparticle, surrounded by metallic Ti$_4$Nb ($Cmcm$) and monoclinic ZrO$_2$ ($P2/c$) phases, confirming the presence of retained metallic regions within the oxide matrix.

Despite the extreme annealing regime with stepwise heating up to 1100~$^{\circ}$C, the Ti–Nb\discretionary{-}{--}{--}Zr–Ag thin film maintained structural integrity, with only local defects such as occasional cracks. The improved structural integrity of the Ag‑containing films could be attributed to the suppression of intergranular oxidation in the presence of Ag. Han et~al.~\cite{Han} showed that the addition of Ag (up to $\sim$15~wt.\%) to Ti significantly reduced oxidation rate and mass uptake at elevated temperatures, suggesting that Ag serves as an oxidation suppressor by altering diffusion pathways and oxide nucleation. This effect is likely related to the tendency of Ag to segregate at grain boundaries, as demonstrated at the atomic scale by Langenohl et~al.~\cite{Langenohl}. Additionally, ductile Ag nanoparticles may contribute to stress relaxation during heat treatment. Unlike bubble-like intralayer cavitation observed in the Ag-free film, which was driven by strong oxidation, the cracks are primarily attributed to the mismatch in volumetric thermal expansion coefficients between the coating ($\sim$26$\cdot 10^{-6}$~K$^{-1}$) and the silicon substrate ($\sim$9$\cdot 10^{-6}$~K$^{-1}$)~\cite{coef}. When applied as a coating on metallic implant materials such as Ti, Ti–6Al–4V, or Ta, which exhibit thermal expansion coefficients closer to that of the coating, the formation of such cracks can be expected to be significantly reduced.

\section{Summary and Conclusions}
\label{sec:summary_and_conclusions}

Ti–Nb–Zr and Ti–Nb–Zr–Ag thin films were deposited by magnetron sputtering and subjected to stepwise annealing up to 1100~°C in order to elucidate the influence of Ag on the structural evolution of medium-entropy alloy coatings. The as-deposited Ag-free coating was fully amorphous, whereas the Ag-containing coating consisted of a predominantly amorphous matrix. Within this matrix were observed dispersed crystalline NbZr, AgZr and $\beta$-(Ti,Nb,Zr) domains as well as Ag nanoparticles, indicating that Ag promotes the onset of crystallization already during deposition. Both coatings exhibited a fine columnar morphology with column widths of approximately 15~nm and dome-like surface protrusions. This surface topography is considered favorable for protein adhesion and thus relevant for potential biomedical applications.

Upon annealing, the two coatings exhibited markedly different structural responses. The Ag-free coating crystallized into a granular and loosely packed microstructure, while the Ag-containing coating largely preserved a compact morphology with only occasional cracking, demonstrating that the presence of Ag enhances the thermal and structural stability of the layer during high-temperature treatment. XPS analysis revealed that annealing led to the formation of both metallic (Ti, Nb) and oxidic species (TiO$_2$, Nb$_2$O$_5$, ZrO$_2$) in both coatings. However, in the Ag-containing coating, lower oxidation states of Ti and Nb were more pronounced after Ar$^+$ sputtering, while Ag remained unoxidized, indicating a reduced degree of deep oxidation of the metallic matrix.

Annealing further induced the formation of ZrO$_2$-like phases with distinct structural characteristics. Nearly stoichiometric regions crystallized in the monoclinic structure, whereas alloyed regions exhibited metastable orthorhombic (Ag-free) or tetragonal (Ag-containing) modifications, often associated with tweed-like contrast arising from chemical and occasionally structural modulation. 

Overall, these results demonstrate that Ag-containing medium-entropy alloy coatings can form dense, amorphous-nanocrystalline composite structures with improved thermal stability. These structures have reduced oxidation susceptibility and biointerface-relevant surface features, while simultaneously benefiting from the well-established antibacterial properties of Ag. We contend that these attributes make such coatings promising candidates for biomedical coating applications.

\section*{Funding}
This work was supported by the MEBioSys project, funded as project No. 
CZ.02.01.01/\-00/22\_008/0004634 by the Johannes Amos Comenius program, call 
Excellent Research. C.D.W.\ acknowledges support from a UK Engineering and Physical Sciences Research Council (EPSRC) Doctoral Prize Fellowship at the University of Bristol, Grant No.~EP/W524414/1.

\section*{CRediT authorship contribution statement}
Anna Benediktová: Conceptualization, Methodology, Investigation, Formal analysis, Visualization, Writing – original draft, Writing – review \& editing.
Lucie Nedvědová: Methodology, Investigation, Writing – original draft, Writing – review \& editing.
Michal Procházka: Investigation, Formal analysis, Visualization, Writing – original draft.
Zdeněk Jansa: Investigation.
Štěpánka Jansová: Investigation.
Christopher D. Woodgate: Conceptualization, Writing – review \& editing.
David S. Redka: Conceptualization, Writing – review \& editing.
Julie Staunton: Conceptualization, Writing – review \& editing.
Ján Minár: Supervision, Conceptualization, Writing – review \& editing.

\section*{Declaration of generative AI and AI-assisted technologies in the manuscript preparation process}
During the preparation of this manuscript, the authors used ChatGPT (OpenAI) to assist in language editing and stylistic refinement of the text. The AI tool was not used to generate scientific content, analyze data, or draw conclusions. The authors reviewed and edited all AI-assisted content and take full responsibility for the content of the published article.

\FloatBarrier
\bibliographystyle{elsarticle-num}
\bibliography{references}

@article{Niinomi,
  author = {Niinomi, M.},
  title = {Recent metallic materials for biomedical applications},
  journal = {Metall Mater Trans A},
  year = {2002},
  volume = {33},
  number = {3},
  pages = {477--486},
  doi = {10.1007/s11661-002-0109-2}
}

@article{Hallab,
  author  = {Hallab, N. J. and Jacobs, J. J.},
  title   = {Biologic effects of implant debris},
  journal = {Bull NYU Hosp Jt Dis},
  year    = {2009},
  volume  = {67},
  number  = {2},
  pages   = {182--188},
  pmid    = {19583551}
}

@article{Jacobs,
  author = {Jacobs, J.J. and Gilbert, J.L. and Urban, R.M.},
  title = {Corrosion of metal orthopedic implants},
  journal = {J Bone Joint Surg Am},
  year = {1998},
  volume = {80},
  number = {2},
  pages = {268--282}
}

@article{Campoccia,
  author = {Campoccia, D. and Montanaro, L. and Arciola, C.R.},
  title = {The significance of infection related to orthopedic devices and issues of antibiotic resistance},
  journal = {Biomaterials},
  year = {2006},
  volume = {27},
  pages = {2331--2339},
  doi = {10.1016/j.biomaterials.2005.11.044}
}

@article{Anselme,
  author = {Anselme, K.},
  title = {Osteoblast adhesion on biomaterials},
  journal = {Biomaterials},
  year = {2000},
  volume = {21},
  pages = {667--681},
  doi = {10.1016/S0142-9612(99)00242-2}
}

@article{Saad,
  author  = {Saad, K. and Saba, T. and Bin Rashid, A.},
  title   = {Application of {PVD} coatings in medical implantology for enhanced performance, biocompatibility, and quality of life},
  journal = {Heliyon},
  year    = {2024},
  volume  = {10},
  number  = {16},
  pages   = {e35541},
  doi     = {10.1016/j.heliyon.2024.e35541}
}

@article{Lian,
  author = {Lian, X. and Cui, H. and Wang, Q. and Song, X. and Yang, X.},
  title = {Micro/nanostructured amorphous {TiNbZr} films to enhance the adhesion strength and corrosion behavior of stainless steel},
  journal = {J Mater Sci Technol},
  year = {2023},
  volume = {164},
  pages = {1--12},
  doi = {10.1016/j.jmst.2023.03.062}
}

@article{Tal,
  author  = {Tallarico, D. A. and Gobbi, A. L. and Paulin Filho, P. I. and Maia da Costa, M. E. H. and Nascente, P. A. P.},
  title   = {Growth and surface characterization of {TiNbZr} thin films deposited by magnetron sputtering for biomedical applications},
  journal = {Materials Science and Engineering: C},
  year    = {2014},
  volume  = {43},
  pages   = {45--49},
  doi     = {10.1016/j.msec.2014.07.013}
}

@article{Gon,
  author  = {Gonzalez, E. David and Fukumasu, Newton K. and Afonso, Conrado R. M. and Nascente, Pedro A. P.},
  title   = {Impact of {Zr} content on the nanostructure, mechanical, and tribological behaviors of {$\beta$}-{Ti}--{Nb}--{Zr} ternary alloy coatings},
  journal = {Thin Solid Films},
  year    = {2021},
  volume  = {721},
  pages   = {138565},
  doi     = {10.1016/j.tsf.2021.138565}
}

@article{Liu,
  author  = {Liu, Chang and Li, Zhiming and Lu, Wenjun and Bao, Yan and Xia, Wenzhen and Wu, Xiaoxiang and Zhao, Huan and Gault, Baptiste and Liu, Chenglong and Herbig, Michael and Fischer, Alfons and Dehm, Gerhard and Wu, Ge and Raabe, Dierk},
  title   = {Reactive wear protection through strong and deformable oxide nanocomposite surfaces},
  journal = {Nature Communications},
  year    = {2021},
  volume  = {12},
  pages   = {5518},
  doi     = {10.1038/s41467-021-25778-y}
}

@article{Panjan,
  author = {Panjan, P. and Drnovšek, A. and Gselman, P. and Čekada, M. and Panjan, M.},
  title = {Review of growth defects in thin films prepared by PVD techniques},
  journal = {Coatings},
  year = {2020},
  volume = {10},
  number = {5},
  pages = {447},
  doi = {10.3390/coatings10050447}
}

@article{Zhang,
  author  = {Zhang, Ergeng and Wang, Yakun and Liang, Dandan and Wei, Xianshun and Zhou, Yinghao and Chen, Qiang and Zhou, Qiong and Huang, Biao and Shen, Jun},
  title   = {Superior corrosion-resistant {Zr--Ti--Ag} thin film metallic glasses as potential biomaterials},
  journal = {Applied Surface Science},
  year    = {2024},
  volume  = {670},
  pages   = {160712},
  date    = {2024-10-15},
  doi     = {10.1016/j.apsusc.2024.160712}
}

@article{Rosalbino,
  author = {Rosalbino, F. and Macciò, D. and Scavino, G.},
  title = {Corrosion behaviour of {Zr--Ag} alloys for dental implant application},
  journal = {Mater Sci Appl},
  year = {2023},
  volume = {14},
  number = {11},
  pages = {369--382},
  doi = {10.4236/msa.2023.1411033}
}

@article{Xu,
  author  = {Xu, Li-Chong and Bauer, James and Siedlecki, Christopher A.},
  title   = {Proteins, Platelets, and Blood Coagulation at Biomaterial Interfaces},
  journal = {Colloids and Surfaces B: Biointerfaces},
  year    = {2014},
  volume  = {124},
  pages   = {49--68},
  date    = {2014-09-28},
  doi     = {10.1016/j.colsurfb.2014.09.040},
  pmid    = {25448722},
  pmcid   = {PMC5001692},
  note    = {A Contribution from the Hematology at Biomaterial Interfaces Research Group}
}

@article{Garcia,
  author  = {Garc{\'\i}a, A. J. and Reyes, C. D.},
  title   = {Bio-adhesive surfaces to promote osteoblast differentiation and bone formation},
  journal = {Journal of Dental Research},
  year    = {2005},
  volume  = {84},
  number  = {5},
  pages   = {407--413},
  month   = {5},
  doi     = {10.1177/154405910508400502},
  pmid    = {15840774}
}

@article{Scop,
  author  = {Scopelliti, Pasquale Emanuele and Borgonovo, Antonio and Indrieri, Marco and Giorgetti, Luca and Bongiorno, Gero and Carbone, Roberta and Podest{\`a}, Alessandro and Milani, Paolo},
  title   = {The Effect of Surface Nanometre-Scale Morphology on Protein Adsorption},
  journal = {PLOS ONE},
  year    = {2010},
  volume  = {5},
  number  = {7},
  pages   = {e11862},
  doi     = {10.1371/journal.pone.0011862}
}

@article{Padamata,
  author  = {Padamata, S. K. and Yasinskiy, A. and Yanov, V. and Saevarsdottir, G.},
  title   = {Magnetron sputtering high-entropy alloy coatings: A mini‑review},
  journal = {Metals},
  year    = {2022},
  volume  = {12},
  number  = {2},
  pages   = {319},
  doi     = {10.3390/met12020319}
}

@article{Anwar,
  author  = {bin Anwar Fadzil, A. F. and Pramanik, A. and Basak, A. K. and Prakash, C. and Shankar, S.},
  title   = {Role of surface quality on biocompatibility of implants -- A review},
  journal = {Ann 3D Print Med},
  year    = {2022},
  volume  = {8},
  pages   = {100082},
  doi     = {10.1016/j.stlm.2022.100082}
}

@article{Chopra,
  author  = {Chopra, Divya and Jayasree, Anjana and Guo, Tianqi and Gulati, Karan and Ivanovski, Sa{\v{s}}o},
  title   = {Advancing dental implants: Bioactive and therapeutic modifications of zirconia},
  journal = {Bioactive Materials},
  year    = {2022},
  volume  = {13},
  pages   = {161--178},
  month   = {7},
  doi     = {10.1016/j.bioactmat.2021.10.010}
}

@article{Savvidou,
  author  = {Savvidou, O. D. and Kaspiris, A. and Trikoupis, I. and Kakouratos, G. and Goumenos, S. and Melissaridou, D. and Papagelopoulos, P. J.},
  title   = {Efficacy of antimicrobial coated orthopaedic implants on the prevention of periprosthetic infections: a systematic review and meta-analysis},
  journal = {J Bone Joint Infect},
  year    = {2020},
  volume  = {5},
  number  = {4},
  pages   = {212--222},
  doi     = {10.7150/jbji.44839}
}

@article{Rashid,
  author  = {Rashid, Saqib and Sebastiani, Marco and Mughal, Muhammad Zeeshan and Daniel, Rostislav and Bemporad, Edoardo},
  title   = {Influence of the Silver Content on Mechanical Properties of {Ti--Cu--Ag} Thin Films},
  journal = {Nanomaterials},
  year    = {2021},
  volume  = {11},
  number  = {2},
  pages   = {435},
  doi     = {10.3390/nano11020435}
}

@article{Yeh,
  author  = {Yeh, J.-W. and Chen, S.-K. and Lin, S.-J. and Gan, J.-Y. and Chin, T.-S. and Shun, T.-T. and Tsau, C.-H. and Chang, S.-Y.},
  title   = {Nanostructured High-Entropy Alloys with Multiple Principal Elements: Novel Alloy Design Concepts and Outcomes},
  journal = {Advanced Engineering Materials},
  year    = {2004},
  volume  = {6},
  number  = {5},
  pages   = {299--303},
  doi     = {10.1002/adem.200300567}
}

@article{Yang2012,
  author  = {Yang, X. and Zhang, Y.},
  title   = {Prediction of high-entropy stabilized solid-solution in multi-component alloys},
  journal = {Materials Chemistry and Physics},
  year    = {2012},
  volume  = {132},
  number  = {2--3},
  pages   = {233--238},
  doi     = {10.1016/j.matchemphys.2011.11.021},
  issn    = {0254-0584}
}

@article{Ma,
  author  = {Ma, Y. and Ma, Y. and Wang, Q. and Schweidler, S. and Botros, M. and Fu, T. and Hahn, H. and Brezesinski, T. and Breitung, B.},
  title   = {High‑entropy energy materials: challenges and new opportunities},
  journal = {Energy Environ Sci},
  year    = {2021},
  volume  = {14},
  pages   = {2883--2905},
  doi     = {10.1039/D1EE00505G}
}

@article{Senkov,
  author  = {Miracle, D. B. and Senkov, O. N.},
  title   = {A critical review of high‑entropy alloys and related concepts},
  journal = {Acta Mater},
  year    = {2017},
  volume  = {122},
  pages   = {448--511},
  doi     = {10.1016/j.actamat.2016.08.081}
}

@article{Zhang2,
  author  = {Zhang, S. and Zhang, Z. and Gao, Y. and Li, H. and Zhang, C. and Zhang, L.},
  title   = {Microstructure and corrosion resistance of a novel AlNiLa lightweight medium‑entropy amorphous alloy composite},
  journal = {J Wuhan Univ Technol Mater Sci Ed},
  year    = {2022},
  volume  = {37},
  pages   = {1185--1191},
  doi     = {10.1007/s11595-022-2651-7}
}

@article{Garah,
  author  = {El Garah, M. and Soubane, D. and Sanchette, F.},
  title   = {Review on mechanical and functional properties of refractory high‑entropy alloy films by magnetron sputtering},
  journal = {Emergent Mater},
  year    = {2024},
  volume  = {7},
  pages   = {77--101},
  doi     = {10.1007/s42247-023-00607-8}
}

@article{Xia,
  author  = {Xia, Q. and Ren, P. and Meng, H.},
  title   = {High performance of amorphous nanocrystalline composite structure materials},
  journal = {J Mater Res Technol},
  year    = {2022},
  volume  = {18},
  pages   = {4479--4485},
  doi     = {10.1016/j.jmrt.2022.04.128}
}

@article{Liu2,
  author  = {Liu, Xinyu and Cai, Wumin and Zhang, Yan and Wang, Linqing and Wang, Junjun},
  title   = {Tuning microstructure and mechanical and wear resistance of {ZrNbTiMo} refractory high-entropy alloy films via sputtering power},
  journal = {Frontiers in Materials},
  year    = {2023},
  volume  = {10},
  pages   = {1145631},
  doi     = {10.3389/fmats.2023.1145631}
}

@article{Feltrin,
  author  = {Feltrin, Ana C. and Xing, Qiuwei and Akinwekomi, Akeem Damilola and Waseem, Owais Ahmed and Akhtar, Farid},
  title   = {Review of Novel High-Entropy Protective Materials: Wear, Irradiation, and Erosion Resistance Properties},
  journal = {Entropy},
  year    = {2023},
  volume  = {25},
  number  = {1},
  pages   = {73},
  doi     = {10.3390/e25010073}
}

@article{Han,
  author    = {Han, X. and Xu, L. and Wu, S. and Wang, G. and Lu, Y.},
  title     = {Microstructure and oxidation behavior of {Ti--Ag} alloys at elevated temperatures},
  journal   = {Materials},
  year      = {2014},
  volume    = {7},
  number    = {9},
  pages     = {6194--6207},
  doi       = {10.3390/ma7096194}
}

@article{Langenohl,
  author  = {Langenohl, Lena and Brink, Tobias and Richter, Gunther and Dehm, Gerhard and Liebscher, Christian H.},
  title   = {Atomic-resolution observations of silver segregation in a {[111]} tilt grain boundary in copper},
  journal = {Physical Review B},
  year    = {2023},
  volume  = {107},
  pages   = {134112},
  doi     = {10.1103/PhysRevB.107.134112}
}

@article{Thornton,
  author  = {Thornton, J. A.},
  title   = {Influence of apparatus geometry and deposition conditions on the structure and topography of thick sputtered coatings},
  journal = {Journal of Vacuum Science and Technology},
  year    = {1974},
  volume  = {11},
  number  = {4},
  pages   = {666--670},
  doi     = {10.1116/1.1312732}
}

@article{Thornton1,
  author  = {Thornton, J. A.},
  title   = {Influence of substrate temperature and deposition rate on structure of thick sputtered Cu coatings},
  journal = {Journal of Vacuum Science and Technology},
  year    = {1975},
  volume  = {12},
  number  = {4},
  pages   = {830--835},
  doi     = {10.1116/1.568682}
}

@article{Amorphous,
  title        = {Amorphous--nanocrystalline alloys: fabrication, properties, and applications},
  author       = {Li, F.-C. and Liu, T. and Zhang, J.-Y. and Shuang, S. and Wang, Q. and Wang, A.-D. and Wang, J.-G. and Yang, Y.},
  journal      = {Materials Today Advances},
  year         = {2019},
  volume       = {4},
  pages        = {100027},
  doi          = {10.1016/j.mtadv.2019.100027}
}

@article{Biesinger2010,
author = {Biesinger, Mark C. and Lau, Leo W.M. and Gerson, Andrea R. and Smart, Roger St C.},
doi = {10.1016/j.apsusc.2010.07.086},
issn = {01694332},
journal = {Applied Surface Science},
number = {3},
pages = {887--898},
publisher = {Elsevier B.V.},
title = {{Resolving surface chemical states in XPS analysis of first row transition metals, oxides and hydroxides: Sc, Ti, V, Cu and Zn}},
volume = {257},
year = {2010}
}

@book{Moulder1995XPSHandbook,
  title     = {Handbook of X-ray Photoelectron Spectroscopy: A Reference Book of Standard Spectra for Identification and Interpretation of XPS Data},
  author    = {Moulder, John F. and Stickle, William F. and Sobol, Peter E. and Bomben, Kenneth D.},
  editor    = {Chastain, Jill and King, Roger C., Jr.},
  year      = {1995},
  publisher = {Physical Electronics, Inc.},
  address   = {Eden Prairie, Minnesota, USA},
  isbn      = {0-9648124-1-X},
  note      = {Copyright 1992, 1995}
}

@article{guo,
  author    = {Guo, S. and Liu, C. T.},
  title     = {Phase stability in high entropy alloys: Formation of solid-solution phase or amorphous phase},
  journal   = {Progress in Natural Science: Materials International},
  year      = {2011},
  volume    = {21},
  number    = {6},
  pages     = {433--446},
  doi       = {10.1016/S1002-0071(12)60080-X}
}

@article{Chua2003,
author = {Chua, Daniel H.C. and Milne, W. I. and Zhao, Z. W. and Tay, B. K. and Lau, S. P. and Carney, T. and White, R. G.},
doi = {10.1016/j.jnoncrysol.2003.09.016},
issn = {00223093},
journal = {Journal of Non-Crystalline Solids},
number = {1--3},
pages = {185--189},
title = {Properties of amorphous {ZrO}$_x$ thin films deposited by filtered cathodic vacuum arc},
volume = {332},
year = {2003}
}

@article{Liao2010,
author = {Liao, Wenchao and Zheng, Tong and Wang, Peng and Tu, Sisi and Pan, Weiqian},
doi = {10.1016/S1001-0742(09)60322-3},
issn = {10010742},
journal = {Journal of Environmental Sciences},
number = {11},
pages = {1800--1806},
pmid = {21235170},
title = {Efficient microwave-assisted photocatalytic degradation of endocrine disruptor dimethyl phthalate over composite catalyst {ZrO}$_x$/{ZnO}},
volume = {22},
year = {2010}
}

@article{Islam2022,
  author  = {Islam, Karimul and Sultana, Rezwana and Satpati, Biswarup and Chakraborty, Supratic},
  title   = {Studies on structural and dielectric properties of {NbO}$_2$--{Nb}$_2${O}$_5$ thin-film-based devices},
  journal = {Vacuum},
  year    = {2022},
  volume  = {195},
  pages   = {110675},
  month   = {1},
  doi     = {10.1016/j.vacuum.2021.110675}
}

@article{Lopez2011,
author = {L{\'{o}}pez, M. F. and Jim{\'{e}}nez, J. A. and Guti{\'{e}}rrez, A.},
doi = {10.1016/j.vacuum.2011.03.006},
issn = {0042207X},
journal = {Vacuum},
number = {12},
pages = {1076--1079},
publisher = {Elsevier Ltd},
title = {XPS characterization of surface modified titanium alloys for use as biomaterials},
volume = {85},
year = {2011}
}

@article{Lee2017,
author = {Lee, Yong Seok and Ryu, Kwang Sun},
doi = {10.1038/s41598-017-16711-9},
issn = {20452322},
journal = {Scientific Reports},
number = {1},
pages = {1--14},
pmid = {29192220},
publisher = {Springer US},
title = {Study of the lithium diffusion properties and high rate performance of {TiNb$_6$O$_{17}$} as an anode in lithium secondary battery},
volume = {7},
year = {2017}
}

@misc{coef,
  author = {{The Engineering ToolBox}},
  title   = {Solids -- Volume Temperature Expansion Coefficients},
  year    = {2025},
  url     = {https://www.engineeringtoolbox.com/volum-expansion-coefficients-solids-d_1894.html},
  urldate = {2025-10-14},
  note    = {Accessed 2025-10-14}
}

@article{Caramarin,
  author  = {Caramarin, Stefania and Badea, Ioana-Cristina and Mosinoiu, Laurentiu-Florin and Mitrica, Dumitru and Serban, Beatrice-Adriana and Vitan, Nicoleta and Cursaru, Laura-Madalina and Pogrebnjak, Alexander},
  title   = {Structural Particularities, Prediction, and Synthesis Methods in High-Entropy Alloys},
  journal = {Applied Sciences},
  year    = {2024},
  volume  = {14},
  number  = {17},
  pages   = {7576},
  doi     = {10.3390/app14177576}
}

@article{Ercan2013,
  author  = {Ercan, Batur and Khang, Dongwoo and Carpenter, Joseph and Webster, Thomas J.},
  title   = {Using mathematical models to understand the effect of nanoscale roughness on protein adsorption for improving medical devices},
  journal = {International Journal of Nanomedicine},
  year    = {2013},
  volume  = {8},
  number  = {Suppl 1},
  pages   = {75--81},
  date    = {2013-09-16},
  doi     = {10.2147/IJN.S47286},
  pmid    = {24098081},
  pmcid   = {PMC3790280}
}

@article{Miedema,
  author = {Miedema, A.R. and de Châtel, P.F. and de Boer, F.R.},
  title = {Cohesion in alloys -- fundamentals of a semi-empirical model},
  journal = {Physica B+C},
  volume = {100},
  number = {1},
  pages = {1--28},
  year = {1980},
  doi = {10.1016/0378-4363(80)90054-6}
}

@article{HallPetch,
  author  = {Cordero, Z. C. and Knight, B. E. and Schuh, C. A.},
  title   = {Six decades of the {Hall--Petch} effect -- a survey of grain-size strengthening studies on pure metals},
  journal = {International Materials Reviews},
  year    = {2016},
  volume  = {61},
  number  = {8},
  pages   = {495--512},
  doi     = {10.1080/09506608.2016.1191808}
}

@article{Bolar,
  author  = {Bolar, Saikat and Ito, Yoshikazu and Fujita, Takeshi},
  title   = {Future prospects of high-entropy alloys as next-generation industrial electrode materials},
  journal = {Chemical Science},
  year    = {2024},
  volume  = {15},
  pages   = {8664--8722},
  doi     = {10.1039/D3SC06784J}
}

@article{Hussein,
  author  = {Hussein, Mohamed Abdrabou and Azeem, Mohammed Abdul and Kumar, Arumugam Madhan and Ankah, Nestor},
  title   = {Design and Development of {Ti--Zr--Nb--Ta--Ag} High Entropy Alloy for Bioimplant Applications},
  journal = {Advanced Engineering Materials},
  year    = {2025},
  volume  = {27},
  number  = {6},
  pages   = {2400462},
  doi     = {10.1002/adem.202400462}
}

@article{Stich,
  author  = {Stich, Theresia and Alagboso, Francisca and Pattappa, Girish and Chu, Jin and Moskal, Denys and Povolny, Michal and Saller, Maximilian and Schoenitzer, Veronika and Scholz, Konstantin J. and Cieplik, Fabian and Alt, Volker and Rudert, Maximilian and Kovarik, Tomas and Krenek, Tomas and Docheva, Denitsa},
  title   = {Micropatterning and Nanodropletting of Titanium by Shifted Surface Laser Texturing Significantly Enhances In Vitro Osteogenesis of Healthy and Osteoporotic Mesenchymal Stromal Cells},
  journal = {Journal of Functional Biomaterials},
  year    = {2025},
  volume  = {16},
  number  = {11},
  pages   = {401},
  doi     = {10.3390/jfb16110401}
}

\end{document}